\documentclass[11pt,letterpaper]{article}
\usepackage[final]{graphics}
\usepackage{graphicx}
\usepackage{makeidx}
\usepackage{authblk} 
\usepackage{epstopdf}
\usepackage{listings}
\usepackage{pstricks,tikz}
\usepackage[absolute,overlay]{textpos}

\usepackage{amsmath, amssymb}
\usepackage{wrapfig}
\usepackage{color,bm}
\usepackage{natbib}
\usepackage[applemac]{inputenc}
\usepackage[portrait]{geometry}

\usepackage{floatrow}
\usepackage{subfig}
\DeclareCaptionLabelFormat{custom}{\fontsize{11}{12}\selectfont(#2)}
\begin{document}

\title{Simulating Moving Contact Lines in Three-Phase Suspensions Using a Front Tracking Method}

\author[1]{Lei Zeng}
\author[1]{Hamideh Rouhanitazangi}
\author[2]{Xianyang Chen}
\author[1]{Jiacai Lu}
\author[1]{Gr\'etar Tryggvason}
\affil[1]{Department of Mechanical Engineering, Johns Hopkins University, MD,  USA}
\affil[2]{Department of Mechanical Engineering, University of Houston, TX,  USA}
\date{October 9, 2024}

\maketitle

\begin{abstract}
Three-phase multiphase flows are found in an extraordinarily large number of applications. Often those involve a liquid phase and a gas phase in addition to a third phase that consists of either liquid drops or solid particles, suspended in the flow. Frequently the third phase is in contact with both the liquid and the gas, resulting in a contact line where all the phases meet. Here, we present an extension of a front tracking method, where the interface between two fluid phases is followed using connected marker points, to simulate the motion of triple contact lines for both three fluids systems and systems containing two fluids and suspended solid particles. We describe two related strategies, one where the contact line is tracked explicitly and one where it is captured implicitly, and show that both approaches achieve comparable accuracy. The second approach is, however, easier to implement, particularly for three-dimensional flows. For both tracked and untracked approaches for solid particles, and for the untracked three fluids case, we use a ``virtual interface,'' where the boundary of a liquid phase is extended into either another fluid or the solid. For three fluids systems the surface tension of the virtual interface is zero, but for systems with solids the surface tension of the virtual interface is the same as that of the physical interface. 
\end{abstract}

\section{Introduction}
In three-phase flows the contact line, where the phases come together, often has an important impact on the evolution. For flows containing only fluids, such as a water-oil-gas mixture the behavior is relatively straightforward, but if one of the phases is a solid the situation is more complex since any motion of the contact line requires a slip at the surface. Here we implement a front-tracking method where a fluid-fluid interface is explicitly tracked by marker points for both a three fluids system as well as for a system consisting of two fluids and suspended solid particles.

Three-phase ``all-fluid'' flows are encountered in, for example, the removal of oil droplets from water (\cite{Saththasivametal2016,Picciolietal2020}). Since the density difference between oil and water is relatively small, gravity separation often takes a long time, sometimes making it impractical. By injecting bubbles into the mixture the rate of separation can be increased. Oil drops are generally hydrophobic and when an air bubble passes by, they hitch a ride and are carried upward, usually spreading over the bubble surface and completely covering it. Other examples include wastewater treatment to remove fat and grease (\cite{Wangetal2010}).

Three-phase flows involving two fluids and solid particles are found in, for example froth flotation, where buoyant bubbles are used to separate hydrophobic particles from a slurry containing both hydrophobic and hydrophilic particles. Froth flotation is the most common mineral separation technique (\cite{King2001}) and is used to separate copper, lead, silver, gold, iron ore, rare earth elements, and coal, from gangue (the commercially uninteresting material), as well as for recycling of plastics (\cite{Fraunholcz2004,Wangetal2015,PitaCastilho2017}), for example. To separate materials, such as metals or coal from mineral matter, the material is ground up, ideally into sufficiently fine particles so that each one consists of nearly pure material. The ground material is suspended in water, forming a slurry. Air is then bubbled through the slurry and when particles collide with the bubbles, hydrophobic particles stick to the bubbles and are carried to the top, where they can be skimmed off, while hydrophilic ones are left behind and sink to the bottom, where they can be removed. While some materials are naturally hydrophobic, such as coal and elemental sulfur, usually the slurry is treated with various chemicals to make the particles that we intend to float hydrophobic, and those that we do not want hydrophilic. The basic principles of froth flotation have been described in a number of books and papers, such as \cite{King2001,FuerstenauHan2003, Fuerstenauetal2007}. 

The dynamics of triple contact junctions (lines in three dimensions and points in two dimensions) where three fluids meet is well described by standard continuum theories. However, if one of the phases is solid in general the contact point moves and we need to augment the standard continuum description to allow a finite slip at the contact point. How to do that is a problem of long-standing interest (\cite{Dussan1979} and \cite{Karim2022}, for example) but here we assume that the slip model can be taken as given. We do, specifically, use a model developed on the basis of molecular simulations by \cite{Qianetal2003,Qianetal2006}, usually referred to as the Generalized Navier Boundary Conditions (GNBC), where the slip velocity is proportional to unbalanced forces at the contact point. The GNBC has previously been used in simulations of stationary solids using front tracking methods by a number of authors, including 
\cite{Yamamotoetal2013,Yamamotoetal2014a,Yamamotoetal2016,Shinetal2018} and \cite{Shangetal2018,Shangetal2019}. For simulations of moving contact lines on stationary solids using other methods, see
\cite{SuiSpelt2013,Afkhamietal2009}, for example.

Several authors have described the integration of a contact line into simulations of three fluids flows.
\cite{Merrimanetal1994, Zhaoetal1996, Smithetal2002} and \cite{Starinshaketal2014} used the level set method; \cite{Li2013} used an arbitrary Lagrangian Eulerian method; and \cite{KimLowengrub2005}, \cite{Kalantarpouretal2020}, and \cite{ZhaoLee} used phase field methods. Other authors have used the moment of fluid method (\cite{Lietal2015c}), and the volume of fluid method, including \cite{PathakRaessi2016} and \cite{Kromeretal2023}, as well as \cite{Zhaoetal2024}, who developed a hybrid height function method for the surface tension. Another recent study can be found in \cite{Garckeetal2024}, where a variational front tracking method is described. 

Although a large number of authors have simulated contact lines moving on a stationary solid, as discussed above, simulations of a moving contact line on a solid that can move freely have only been done by a few authors. \cite{Fujitaetal2013, Fujitaetal2015} used a level set method, \cite{Pateletal2017} and \cite{OBrienBussmann2020} employed a volume of fluid method and immersed boundary method, and \cite{Nguyenetal2021} used a discrete element method.

\section{Formulation}
The motion of the fluid is taken to be governed by the Navier-Stokes equations and incompressible. In conservative form, using the one-fluid formulation, the governing equations therefore are:
\begin{equation}
\frac{\partial \rho {\bf u}}{\partial t}+ \nabla \cdot \rho {\bf u}{\bf u}=-\nabla p + \rho {\bf g}+\nabla \cdot {\bm \tau} +{\bf f}
\quad \hbox{and} \quad \nabla \cdot {\bf u}=0.
\label{navierstokes}
\end{equation}
Here, ${\bf u}$ is the velocity vector, $\rho$ is the discontinuous density field, $p$ is the pressure, and ${\bf g}=-g{\bf k}$, where $g$ is the gravity acceleration and ${\bf k}$ is a unit vector in the vertical direction.
The viscous stresses in the fluids are given by ${\bm \tau} =2 \mu {\bf D}$, where ${\bf D}=(1/2) (\nabla {\bf u}+ \nabla {\bf u}^T)$ is the deformation tensor and $\mu$ is the viscosity. For a solid, the stresses are whatever is required to keep it rigid. ${\bf f}$ is a force that that represents surface tension for fluid-fluid interfaces (${\bf f}_{\sigma}$) and enforces rigidity of solid particles (${\bf f}_{s}$). For solid particles it also includes a collision force to keep the particles from overlapping.
The different phases are identified by indicator functions
\begin{equation}
\chi_i=  \left\{ 
  \begin{array}{l l}    
1 \quad &  \text{in phase $i$}\\
0  \quad &  \text{everywhere else}.
  \end{array} \right.
\end{equation}
The indicator function moves with the fluid velocity
\begin{equation}\frac{\partial \chi_i}{\partial t} + {\bf u} \cdot \nabla \chi_i =0,
\end{equation}
and the material properties are found from the indicator functions. 

\begin{figure}
\centering{
\sidesubfloat[]{\label{Sketch1:a}
    \includegraphics[scale=1.0]{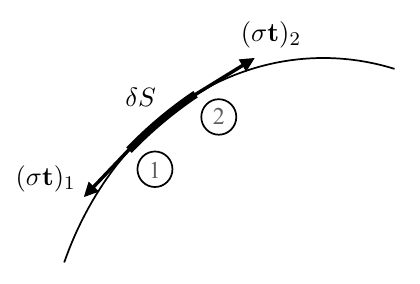} }
\ \ \
\sidesubfloat[]{\label{Sketch1:b}
    \includegraphics[scale=1.0]{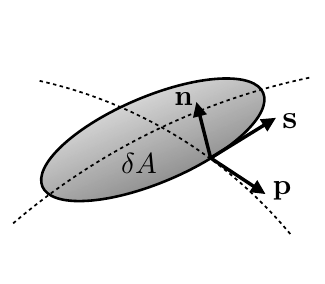} }}
\caption{An interface in (a) two dimensions and (b) three dimensions.}
\label{Sketch1}
\end{figure}

For two-dimensional flow, where the interface is a line, the interface tension is given by $\sigma {\bf t}$, where $ {\bf t}$ is tangent to the interface. Thus, the net force on the interface segment shown by the thick line in Figure~\ref{Sketch1}\subref{Sketch1:a} is given by the difference in the tension at points 1 and 2, pulling to the left and the right, or by
\begin{equation}
{\bf f}_{\sigma}= \int_{\delta S} \frac{\partial \sigma {\bf t}}{\partial s} ds = (\sigma {\bf t})_2 -(\sigma {\bf t})_1.
\label{2Dsigma1}
\end{equation}
In the limit of constant surface tension and zero segment length, this reduces to the product of the curvature and the normal to the interface. For three-dimensional flow, where the interface is a surface, the force on the small segment shown in gray in Figure~\ref{Sketch1}\subref{Sketch1:b} is given by the integral around the edges 
\begin{equation}
{\bf f}_{\sigma}= \oint \sigma  {\bf s} \times {\bf n} ds .
\label{3Dsigma1}
\end{equation}
Here, ${\bf s} $ is a tangent vector to the segment boundary, ${\bf n}$ is a normal to the surface, and the vector ${\bf p} ={\bf s} \times {\bf n}$ ``pulls'' on the edge and, again, the integral reduces to the product of the mean curvature and the normal for constant surface tension and in the limit of zero segment size. Here we do not assume constant surface tension and therefore work directly with equations (\ref{2Dsigma1}) and (\ref{3Dsigma1}).

\subsection{Moving Triple Contact Line for Three Fluids}
\begin{figure}
\centering
\sidesubfloat[]{\label{TripleLine1:a}
    \includegraphics[scale=0.14]{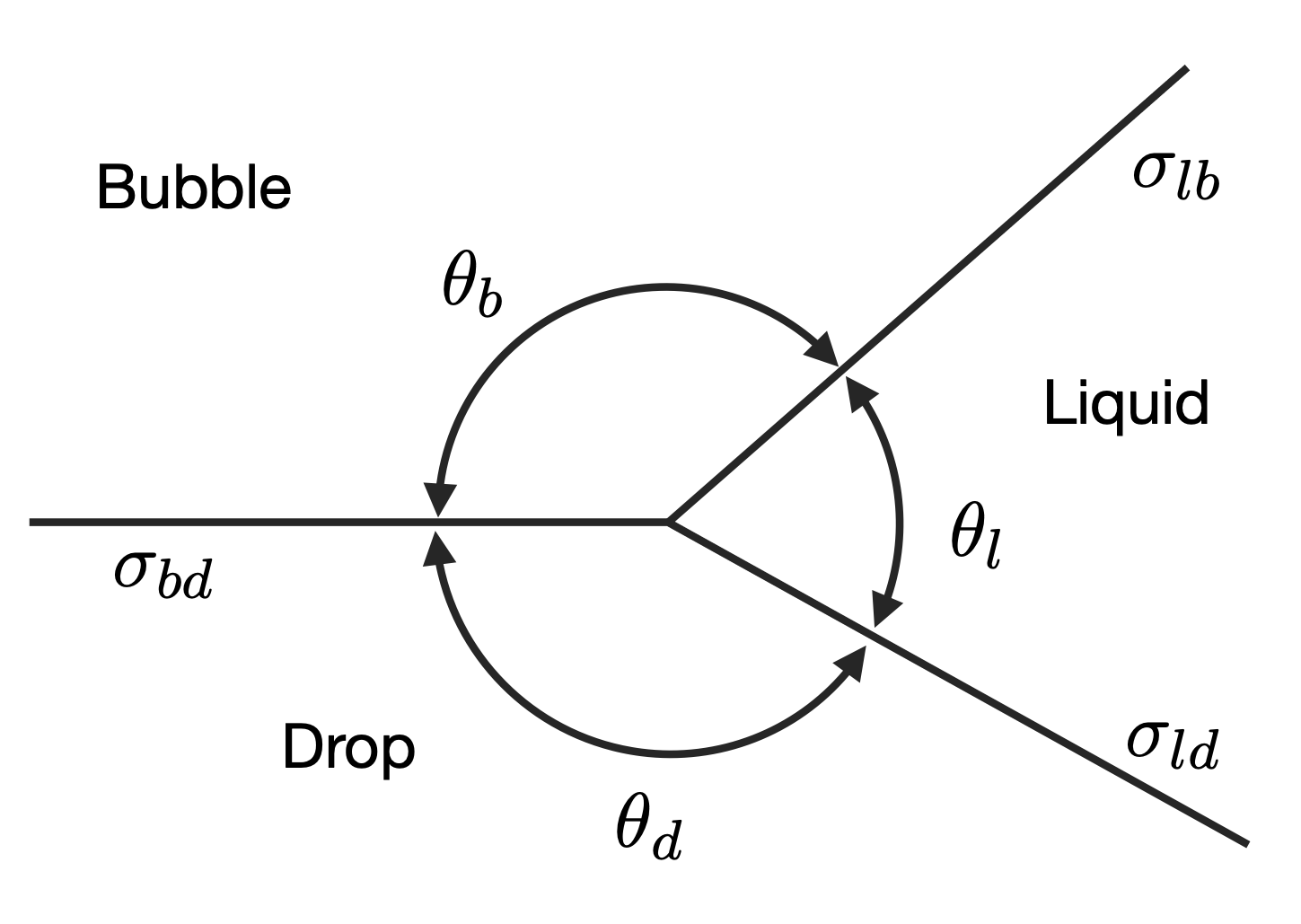}}
\sidesubfloat[]{\label{TripleLine1:b}
    \includegraphics[scale=0.14]{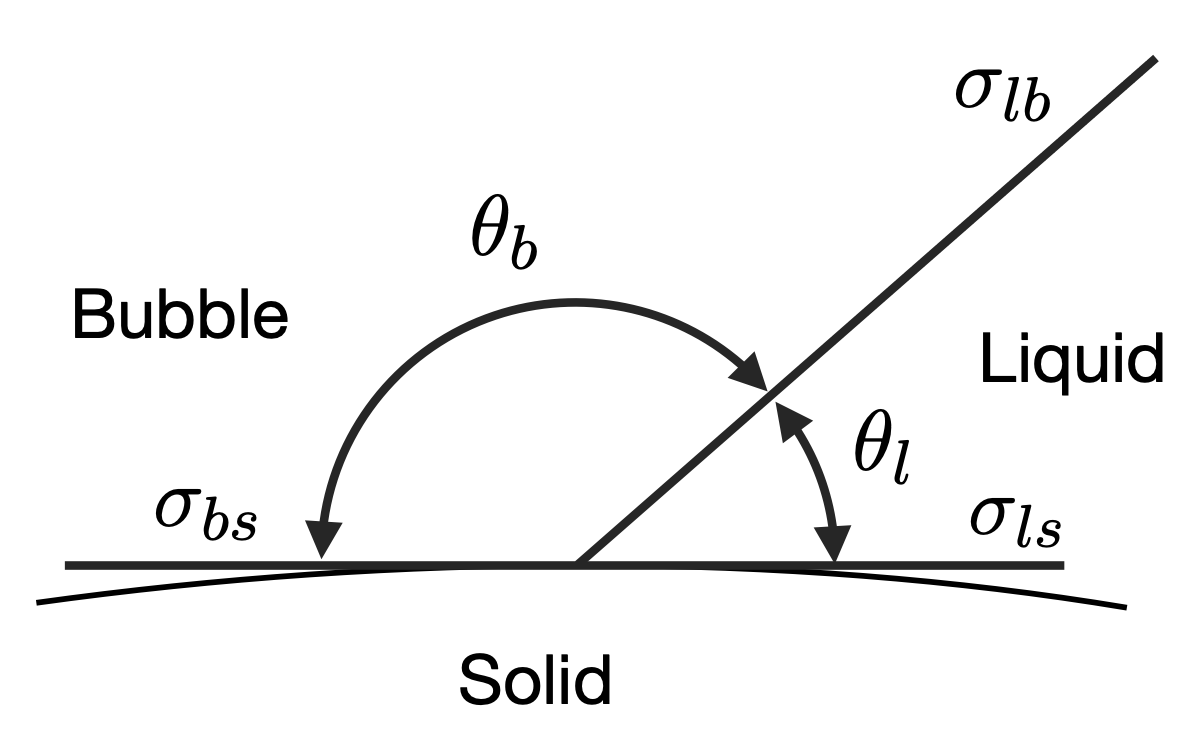}}
\caption{The notation for (a) three fluids and (b) two fluids and a solid.}
\label{TripleLine1}
\end{figure}

If each phase is only in contact with one other phase, the formulation above (equations (\ref{navierstokes}-\ref{2Dsigma1} or \ref{3Dsigma1}) is complete. However, if three phases meet at a triple point (two dimensions) or triple line (three dimensions) additional considerations are needed. Figure~\ref{TripleLine1}\subref{TripleLine1:a} shows a triple point where three fluids (which we will refer to as the liquid, the bubble, and the drop) meet.
To do a force balance, we put an arbitrarily small control volume around the triple point. The different interfaces pull on the triple point, exerting a net force of
\begin{equation} 
{\bf f}_{\sigma} =\sigma_{lb} {\bf t}_{lb} + \sigma_{bd} {\bf t}_{bd} + \sigma_{ld} {\bf t}_{ld},
\label{fluidforces1}
\end{equation} 
which needs to be added to the fluid equations.
The horizontal and vertical components of the force (using the orientation in the figure) are
\begin{equation} 
f_h=\sigma_{bd} + \sigma_{lb} \cos \theta_b +\sigma_{ld} \cos \theta_d,
\label{hbalance}
\end{equation} 
\begin{equation} 
f_v=\sigma_{lb} \sin \theta_b - \sigma_{ld} \sin \theta_d.
\label{vbalance}
 \end{equation} 
If there is no motion, the only forces are the interface tensions, so at equilibrium the sum of the forces must be zero, $f_h=f_v =0$, and we can solve for the equilibrium angles (see Appendix 1): 
\begin{equation} 
 \cos \theta_l=\frac{1}{2} \Bigr( \frac{\sigma_{bd}}{\sigma_{ld}} \frac{\sigma_{bd}}{\sigma_{lb}}
-\frac{\sigma_{ld}}{\sigma_{lb}} -\frac{\sigma_{lb}}{\sigma_{ld}} \Bigr),
 \quad 
\cos \theta_b=\frac{1}{2} \Bigr( \frac{\sigma_{ld}}{\sigma_{bd}} \frac{\sigma_{ld}}{\sigma_{lb}}
-\frac{\sigma_{bd}}{\sigma_{lb}} -\frac{\sigma_{lb}}{\sigma_{bd}} \Bigr).
\label{generalsolution1}
\end{equation}
The third angle can be found by using the fact that the angles must add up to $2 \pi$.

We note that steady state solutions involving a triple line are only realized for certain combinations of surface tensions. From the figure, we see that there is no equilibrium solution if $\sigma_{bd}> \sigma_{ld}+ \sigma_{lb}$, since even for $\theta_l=0$ there will be a stronger pull to the left than to the right. Thus, the largest surface tension must be smaller than the sum of the other two, and if the surface tension between two of the phases is zero, a steady state with a triple line is only possible when the other two surface tensions are equal.
If steady state with a triple line is not possible, the drop coats the bubble if $\sigma_{lb} > \sigma_{ld}$  and if $\sigma_{lb} < \sigma_{ld}$ the drop is engulfed by the bubble. In general, if a steady state solution is not possible, it means that either the bubble or the drop is engulfed by the other, or they have separated.
For oil droplets and air bubbles in water, for example, we can take $\sigma_{lb}=72 mN/m$, $\sigma_{ld}=25 mN/m$ and $\sigma_{bd}=30 mN/m$, which means that the air bubble is engulfed into the oil drop since $\sigma_{lb}> \sigma_{ld}+\sigma_{bd}$. 
It is also clear that if two of the surface tensions are equal, say $\sigma_{lb}=\sigma_{ld} $  then equations (\ref{generalsolution1}) tells us that the angles are equal and we have $ \cos \theta_b= \cos \theta_d= -(1/2) {\sigma_{bd} / \sigma_{lb} }$. Similarly, if $\theta_b=\theta_d=\pi/2$, then we must have $\sigma_{bd}=0$. 
 
The angles here are the equilibrium angles and when they are different, there is a net force at the triple point that is balanced by other fluid forces such as viscous stresses. However, in the absence of other forces, the system will move toward the equilibrium angles, and equations (\ref{generalsolution1}) can be used for verification of the results.

\subsection{Moving Triple Contact Line for an Embedded Solid}
In principle, the considerations for the equilibrium contact angle are similar when one of the fluids is replaced by a solid. For the discussion here, we assume that the drop is replaced (Figure~\ref{TripleLine1}\subref{TripleLine1:b}). At steady state the interface forces must balance. Unlike the all-fluid case, here a smooth solid surface remains locally flat at the contact point as is seen in the sketch, where it is clear that the angle in the solid is $\theta_s=\pi$. We can do a force balance as before and find that the forces tangent and normal to the solid surface are given by
\begin{equation}
f_t = \sigma_{ls}-\sigma_{bs} +\sigma \cos \theta_l,
\label{solidtangent1}
\end{equation}
\begin{equation}
f_n = \sigma  \sin \theta_l,
\label{solidnormal1}
\end{equation}
where we have left out the subscript on the surface tension for the gas-liquid interface.
If there is a net force so that $f_t \ne 0$, the interface will generally move, but if it is stationary we find that 
\begin{equation}
\sigma_{ls}-\sigma_{bs} =-\sigma \cos \theta_o,
\label{solidtangent2}
\end{equation}
where $\theta_o$ is the equilibrium contact angle.
Thus, we can rewrite the tangent force by substituting equation (\ref{solidtangent2}) into (\ref{solidtangent1}), yielding
\begin{equation}
f_t = \sigma  ( \cos \theta_l - \cos \theta_o).
\label{solidtang1}
\end{equation}

When the contact angle is different from the equilibrium angle and there is a net tangent force at the contact point, it will move with what is usually called the slip velocity, since the solid surface is rigid. A slip velocity at a solid surface leads to a force singularity and many models have been proposed to predict the slip as a function of the force and the properties of the fluids and the solid. 
In the present study we use the Generalized Navier Boundary Conditions (GNBC) model, which states that the slip is proportional to any unbalanced force
\begin{equation}
\beta \Delta u_{slip} = \tau_{visc} - f_t .
\label{slipvel1}
\end{equation}
Here, $ \beta=\mu / \lambda$, where $\lambda$ is a slip length, $\tau_{visc}$ represents viscous forces, and $f_t$ is the force due to surface tension,  given by equation (\ref{solidtang1}). We denote the slip velocity by $\Delta u$ to emphasize that it is the slip velocity in a frame moving with the solid. The viscous forces are often ignored and we will do so here.  For our purpose, the slip length is simply an empirical proportionality coefficient relating the slip to the force so we write 
\begin{equation}
\Delta u_{slip} = C_{slip} f_t.
\label{slipvel2}
\end{equation}

For both drops and solid particles we should also specify under what conditions a contact line is formed when the drop or solid is attached to the bubble, and how detachment takes place. However, as discussed in the next section on the numerical implementation, we are currently using a very simple model and will thus leave out a detailed description here. The attachment depends on the specific conditions at the interface and a more detailed discussion can be found in, for example, \cite{Huangetal2005, Wangetal2016, Xingetal2017}.

\begin{figure}
\centering
\sidesubfloat[]{\label{TwoMethods:a}
    \includegraphics[scale=1]{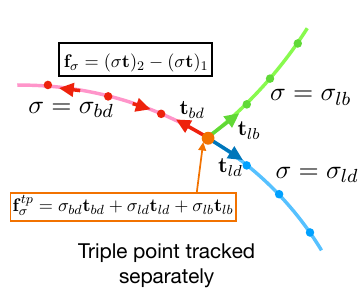}}
\ \ \
\sidesubfloat[]{\label{TwoMethods:b}
    \includegraphics[scale=1]{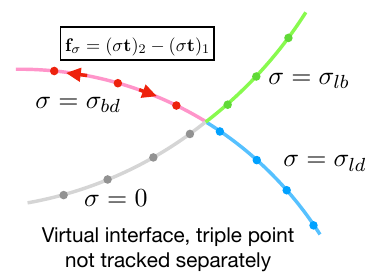}}
\caption{A sketch showing the different strategies implemented to handle triple points for three fluids problems. (a) The triple point is treated separately. (b) The surface tension changes abruptly at the triple point but the point is not explicitly tracked.}
\label{TwoMethods}
\end{figure}

\section{Numerical Approach}
We have implemented a moving contact line into the same flow solver for both the all-fluid case and when one of the phases is solid. In both cases we have modeled the moving contact line in two different ways, one where it is explicitly tracked by a separate point (in two dimensions) and where it is not. We first describe the flow solver briefly before discussing the incorporation of the contact line.

\subsection{Flow Solver}
The Navier-Stokes equations are solved on a regular structured staggered grid using a finite volume method that is second order in space and time. The method has been implemented both as a simple Matlab code for two-dimensional flow and as a fully parallelized Fortran code for three-dimensional flow. The diffusion terms are discretized using centered differences and the advection terms are discretized using second order QUICK upwind scheme for three-dimensional flows and centered differences for two-dimensional flows. The non-separable pressure equation is solved by Hypre (\cite{FalgoutYang2002}) in three dimensions and a simple SOR method in two dimensions.

To track a moving fluid-fluid interface we use connected marker points, as shown in Figure~\ref{TwoMethods}, that move with the fluid velocity, interpolated from the fixed grid. The connected marker points are used to construct indicator functions that identify the different fluids/phases and to find surface tension. The velocity of the marker points is interpolated from the fixed fluid grid and surface forces, found at the front marker points, are smoothed onto the fixed fluid grid.
This approach is usually referred to as front tracking and has been described in detail in \cite{TSZ:2011}, for example. 

For solids the surface tension is zero, so the only role of the marker points (the front) is to allow the construction of the indicator function. However, if the solid has a relatively simple shape that can be described analytically, such as a spheroid, the indicator function can be constructed directly from the analytical expression. Thus, we use a front to construct the indicator function for bubbles and drops, but not for the solids. We note that for solids we generally find it useful to replace the indicator function $\chi_s$ by its square $\tilde \chi_s=\chi_s^2$ for faster convergence (\cite{Luetal2023}).

For solids, we need to enforce zero deformation. This can be done in many ways and here we treat the solid as a fluid region but add a force to keep it rigid. We first assign a high viscosity to the solid region, and then advect the indicator function and update the velocities everywhere, including inside the solid, in the same way as done for an all-fluid problem. Then we compute the centroid velocity, ${\bf u}_c $, and rotation rate, ${\bf\Omega}_s$, for the solid particle by integrating over it:
\begin{equation}
{\bm u}_c = \frac{\int \tilde \chi_s \rho_s {\bm u} dv}{\int \tilde \chi_s \rho_s dv}; \qquad
{\bm \Omega}_s = \frac{\int \tilde \chi_s {\bm r} \times \rho_s {\bm u}  dv}{\int \tilde \chi_s \rho_s ({\bm r} \cdot {\bm r}) dv},
\end{equation}
where ${\bf r}$ is the location of a point inside the solid with respect to its centroid. In general, the velocity inside the solid is not equal to the velocity field computed from the centroid velocity and the rotation rate and we add an extra step where we update the velocity by adding a force
\begin{equation}
{\bf f}_s = \frac{1}{\Delta t} ({\bf u}_s -{\bf u})\quad \hbox{where} \quad
{\bf u}_s={\bf u}_c + {\bf\Omega}_s \times {\bf r}.
\end{equation}
If the artificial viscosity inside the solid is high, the deformation and velocity correction during one time step are very small.

In general, we expect the indicator functions to be mutually exclusive since only one phase can occupy each spatial location, but because we modify the indicator functions for computational purpose, allowing slight overlap, we need to construct the material properties, such as the density, from the indicator function sequentially: First the properties are set equal to that of the liquid; then using the indicator function for the bubble to override the values for the liquid inside the bubble; and finally the properties are set using the indicator function inside the drops, overriding the values for the liquid and the bubble. For example, for the densities, the steps are:
\begin{eqnarray}
\hbox{1: } \rho&=\rho_l \ \ \ \ \\
\hbox{2: } \rho&=\chi_g \rho_g  \\
\hbox{3: } \rho&=\chi_{d} \rho_{d}.
\end{eqnarray}
Other properties are set in the same way. 

We have not attempted to model attachment and detachment in any detail yet. Currently we simply attach a solid or a drop that collides with the bubble by moving one element of the front representing the bubble interface onto the solid/drop, if the bubble and the solid/drop are moving towards each other. We also monitor the relatively tangent velocities of the bubble and the solid/drop and only reconnect if it is small. Our detachment model is also very simple. Once the area shared by the bubble and the solid/drop shrinks below a prescribed threshold, we simply move the bubble interface slightly out of the solid/drop. In reality the attachment is more complex. The film between the bubble and the solid/drop has to have time to drain and that can depend on the detailed surface conditions of the bubble surface, in addition to whether a solid is hydrophobic or hydrophilic. We believe that more sophisticated models can be added, although that will include additional modeling and code development.

\begin{figure}
\centering
{\includegraphics[scale=0.75]{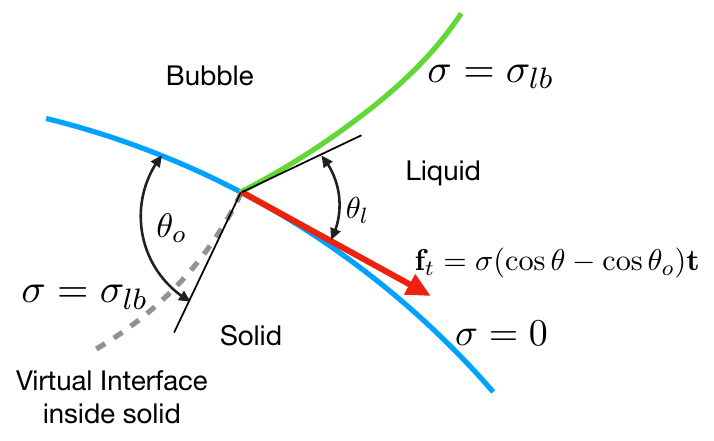} }
\caption{Balancing the forces at the solid surface using a virtual interface inside the solid. At equilibrium $\theta_l=\theta_o$ and $f_t=0$.}
\label{VirtualInterface1}
\end{figure}

\subsection{Three Fluids}
For the three fluids problem, we have implemented two different strategies to follow the contact point, shown in Figure~\ref{TwoMethods}. The first method involves explicitly tracking the triple point or line and computing the surface force at the triple point by equation (\ref{fluidforces1}), smooth it onto the fixed grid and add it to the Navier-Stokes equations in exactly the same way as the surface force for regular interface markers.

Tracking the triple point separately adds considerable code complexity, particularly in three dimensions, and motivates the second approach, shown in Figure~\ref{TwoMethods}\subref{TwoMethods:b}. Here we keep each bubble, drop, or solid particle intact and enclosed by a closed surface by continuing one interface inside another phase.
We refer to this continuation as a virtual interface. The red, green and blue lines are the physical interfaces (the same as Figure~\ref{TwoMethods}\subref{TwoMethods:a}), but the gray line represents a continuation of the bubble interface through the drop. The surface tension of the drop interface inside the bubble is set to $\sigma_{bd}$, but the gray interface has zero surface tension. To identify what part of an interface is the virtual one, we generate an interface indicator function $I(s)$ which is zero everywhere except when it crosses another disperse phase. Thus, for the bubble interface we interpolate the indicator function of the drop, and for the drop we interpolate the bubble indicator function. For three fluids flows nothing else needs to be done, all surface forces are found using equation (\ref{2Dsigma1}), and the force balance at the triple point is straightforward since the surface tensions are known. In principle, the shape of the virtual interface is of no importance since its surface tension is zero, and the simplest approach is to move it with the fluid velocity, as the other interfaces. However, this can result in it coming very close to the bubble/drop interface so the interpolation of the indicator function becomes inaccurate. Thus, we modify its location slightly to keep a smooth continuation, in a similar way as done for a solid and discussed in the next section. We also note that we only need the virtual interface to extend slightly into the solid, since the rest is completely passive, but using a closed surface simplifies the maintenance of the front---at least in our implementation of the method.

\subsection{Two Fluids and a Solid}
When we replace the drop with a solid, the force in the direction normal to the surface is balanced by forces in the solid that we do not know and should be found, in principle, by imposing forces in the solid to keep it rigid (assuming a completely undeformable solid). For a stationary solid this is not a major concern because the rigidity of the solid is enforced regardless of whether the forces are exactly balanced. For a solid that is free to move it is, on the other hand, critical to implement the normal force in a way that conserves momentum. We have implemented an interface pinned to a solid surface but generally find that the deformation is confined to a small volume and it is difficult to ensure complete momentum conservation. The use of a virtual interface, where the interface separating the bubble and the liquid is continued into the solid, simplifies how we account for the force in the solid. This approach was introduced by \cite{Fujitaetal2013, Fujitaetal2015} in simulations of floating particles using a level set method, and has been used by \cite{Nguyenetal2021}, for example. If the surface tension of this virtual interface is the same as the real interface, and the interface in the solid has the same tangent vector at the surface, then it is easily shown that the virtual interface provides the necessary solid force at equilibrium.

The virtual interface connects all points at the contact line so in two dimensions it connects two points where the real interface ``enters'' and ``exists'' the solid. The situation is sketched in Figure~\ref{VirtualInterface1}. At equilibrium, the interface enters the solid smoothly and $\theta_l=\theta_o$, but when the angle between the interface and the solid is not the equilibrium angle, then there is a discontinuity in the slope since the angle inside the solid must be the equilibrium angle. Doing a force balance at the contact point gives equations (\ref{solidtangent1}) and (\ref{solidnormal1}). 
The normal force is given by equation (\ref{solidnormal1}) and the tangent by equation (\ref{solidtang1}). 
The tangent to the virtual interface is determined by the equilibrium angle and this determines the location of the first interface points inside the solid. It seems most reasonable to keep the force in the solid constant by keeping the curvature as constant as possible.
To enforce a constant curvature, we set the coordinates of the first point inside the solid based on the equilibrium angle and then distributed points on a smooth curve connecting those points. The location of the points is  adjusted iteratively by minimizing a discrete version of
\begin{equation}
E=\int \Big\vert \frac{\partial^2 {\bf x}}{\partial s^2} \Big\vert^2 ds.
\end{equation}
The details of the iterative scheme are given in Appendix 2. A similar approach is used to keep the virtual interface smooth inside the drop in the all-fluid problem.

When we track the contact point, the tangent force at that point is easily computed by equation (\ref{solidtang1}), and the slip tangent velocity is found by equation (\ref{slipvel2}). 
For the untracked approach we use the closest interface point as the contact point, and as the grid is refined it converges to the physical contact point. The contact points in both approaches are moved with the solid velocity plus the slip velocity. Points in the fluid, away from the solid, move with the interpolated fluid velocity and to ensure a smooth transition in the interface velocity, we linearly blend the fluid velocity and the velocity of the contact point for interface points within three grid spacings from the solid.

\section{Results}

\subsection{Results for Three Fluids}

\begin{figure}
\centering{
\sidesubfloat[]{\label{dropresult:a}
    \includegraphics[scale=0.275]{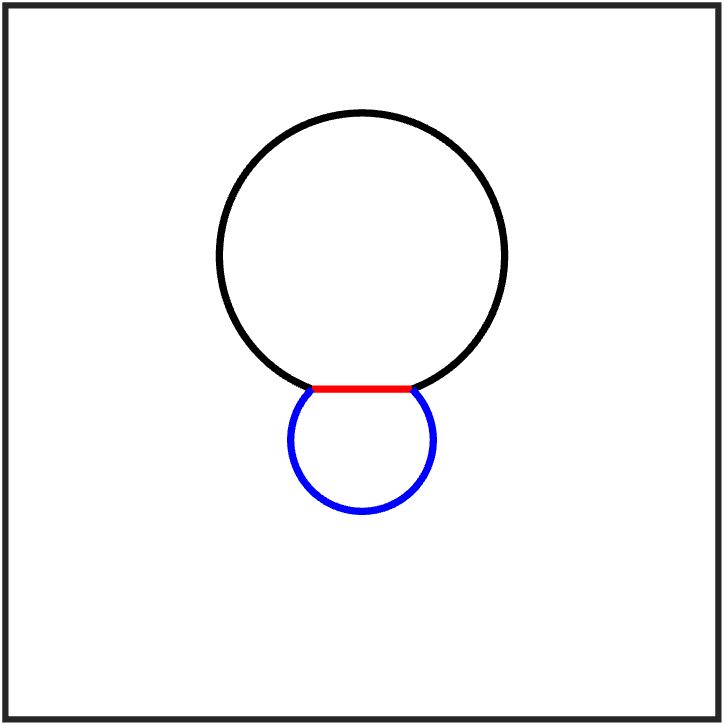}
    \includegraphics[scale=0.275]{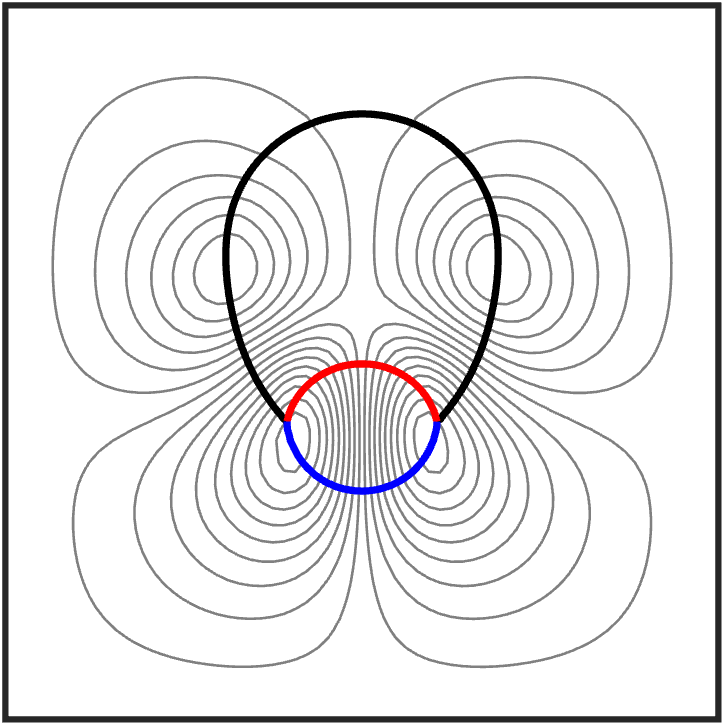}
    \includegraphics[scale=0.275]{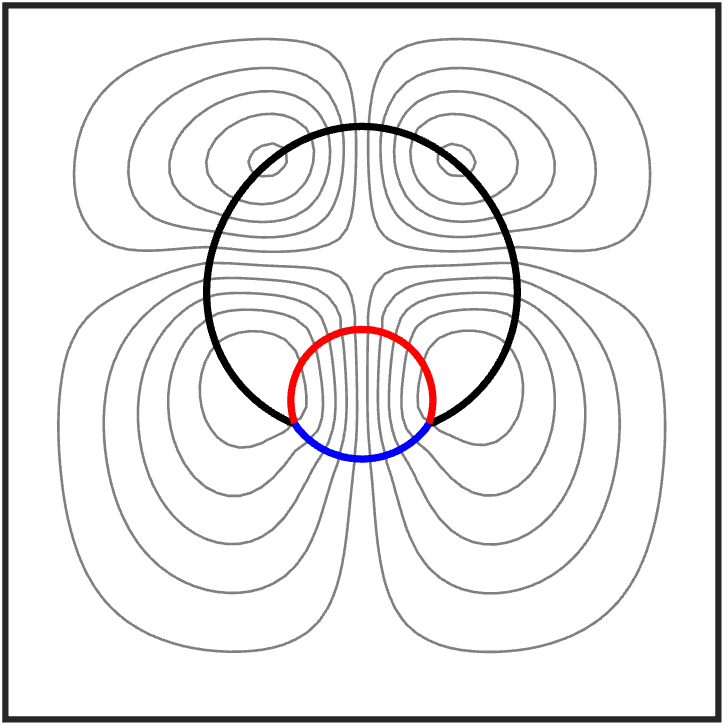}
    \includegraphics[scale=0.275]{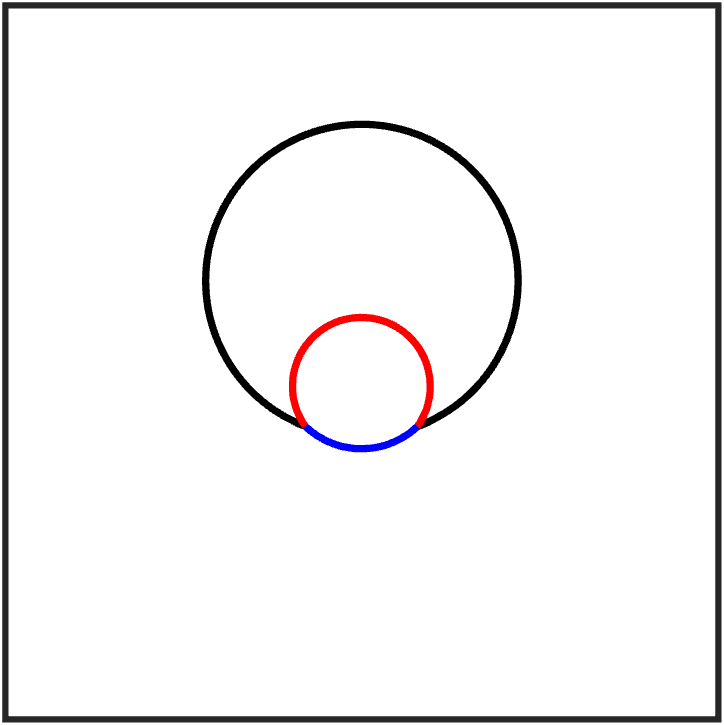}}
\vspace{.2cm}
\sidesubfloat[]{\label{dropresult:b}
    \includegraphics[scale=0.275]{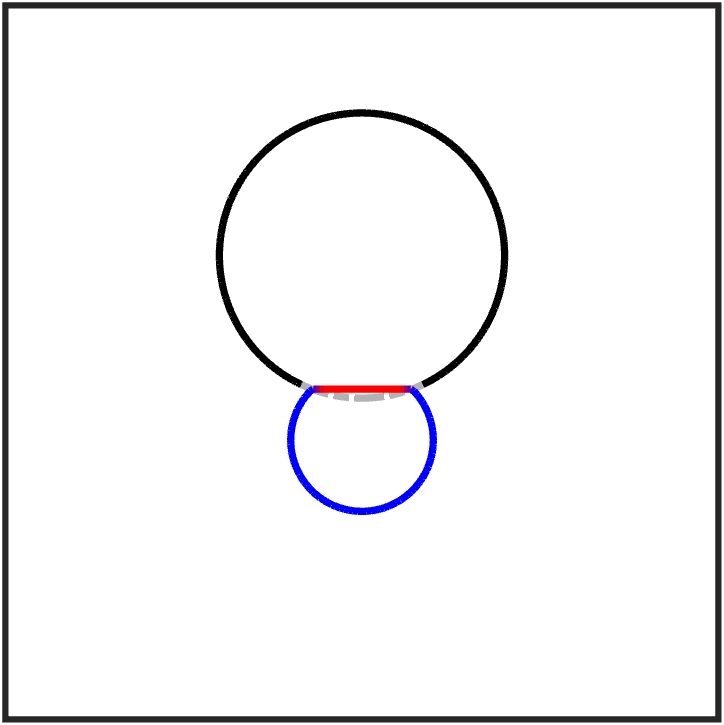}
    \includegraphics[scale=0.275]{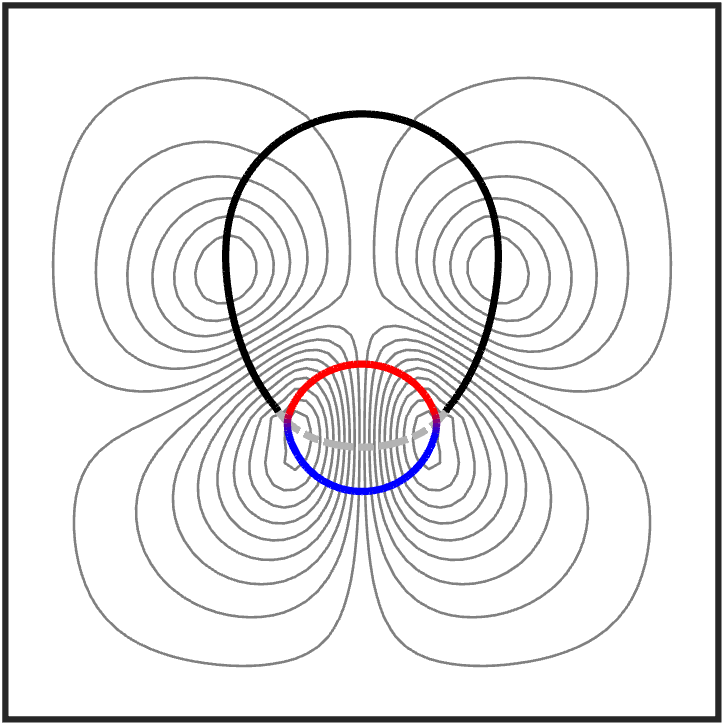}
    \includegraphics[scale=0.275]{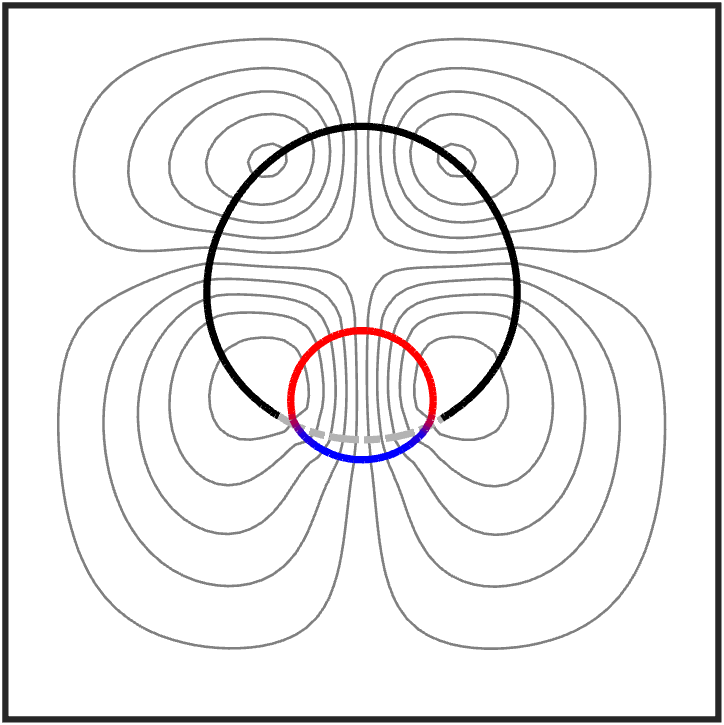}
    \includegraphics[scale=0.275]{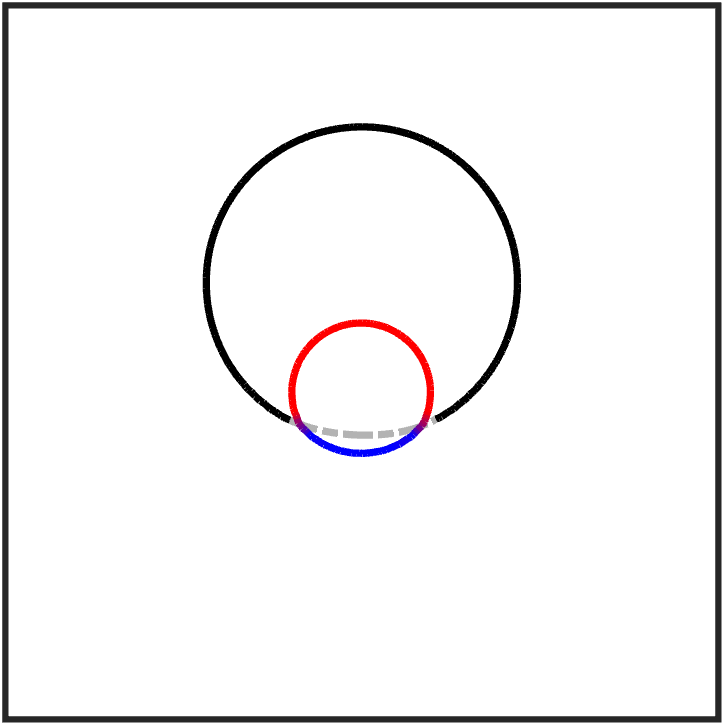}}}
\caption{The transient evolution for two-dimensional flow, computed in two different ways. (a) The triple points are explicitly tracked. (b) The triple points are not tracked but the bubble surface is extended into the drop as a virtual interface with zero surface tension.
Times are, from left to right,  0.0, 0.2, 0.5, 2.0.}
\label{dropresult}
\end{figure}

\begin{figure}
\centering{
\includegraphics[scale=0.45]{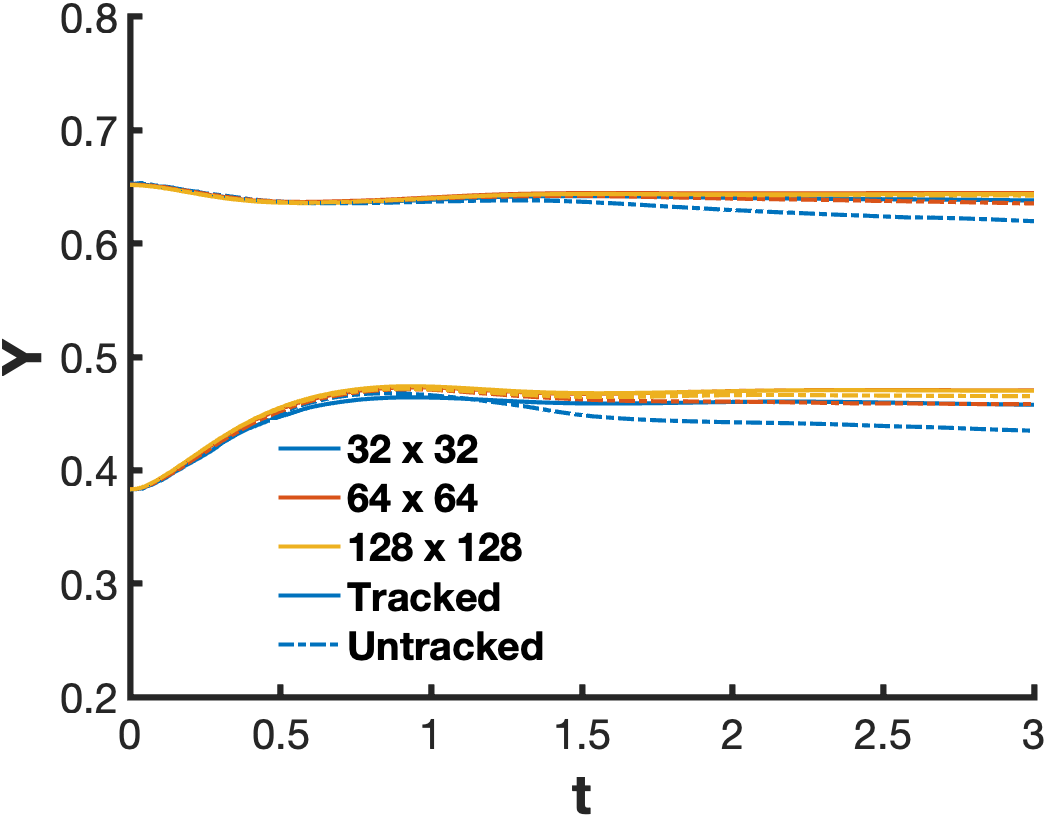}}
\caption{The motion of the centroids of the bubble and drop versus time for three different resolutions and both approaches.}
\label{centroid_resolution_test}
\end{figure}

We have implemented both methods into a simple Matlab code for two-dimensional flows and in Figure~\ref{dropresult} we show the transient evolution for a drop of diameter $d_{d} =0.2$ attached to a bubble of $d_{b} =0.4$ in a $1 \times 1$ domain resolved by $64 \times 64$ regular grid. The evolution in Figure~\ref{dropresult}\subref{dropresult:a} is computed by tracking the triple point and in Figure~\ref{dropresult}\subref{dropresult:b} we use the second method where we do not track it but instead continue the bubble interface through the drop as a virtual interface (shown as a gray dashed line because surface tension is zero). The densities are $\rho_{l}=1.0$, $\rho_{b}=0.5$, and $\rho_{d}=1.5$, and the viscosities are $\mu_{l}=0.02$, $\mu_{b}=0.01$, and  $\mu_{d}=0.05$. The surface tension coefficients are $\sigma_{lb} =0.1$ between the liquid and the bubble; $\sigma_{ld} =0.29 $ between the liquid and the drop; and $\sigma_{bd} =0.2$ between the bubble and the drop. These values result in an equilibrium contact angle of $\theta_{l}=158.8^o$ and an Ohnsorge number of $Oh_d=\mu_l / \sqrt{\rho_l \sigma_{lb} d_d}=0.141$ for the drop. The drop is initially placed so the interfaces overlap (first frame) and the drop is then pulled into the bubble until reaching an equilibrium (last frame). In addition to showing the fluid interfaces, we plot a few streamlines for the times when the fluids are moving. The location of the centroids of the bubble and the drop are shown in Figure~\ref{centroid_resolution_test}. The results are essentially the same for both methods and all resolutions, except that for the lowest resolution the results from the untracked approach show a slight downward drift of both the bubble and the drop at a later time.

\begin{figure}
\centering{\includegraphics[scale=1.0]{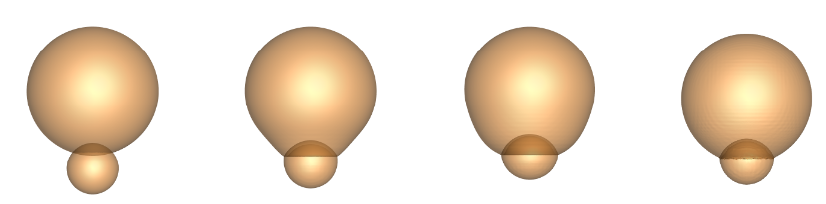}}
\caption{The transient evolution for a fully three-dimensional flow. Times are, from left to right, 0.0, 0.2, 0.6, 2.0.}
\label{Results3D}
\end{figure}

A similar evolution is shown in Figure~\ref{Results3D} for fully three-dimensional flow, computed using the untracked approach. The initial conditions are shown in the first frame and the steady state solution in the last frame. Here, the densities and viscosities are the same for all the fluids $\rho_l=\rho_b=\rho_d=1.0$ and $\mu_l=\mu_b=\mu_d=0.01$. The bubble diameter is $d_b=0.5$ and the drop diameter is $d_d =0.2$. The computations are done in a $1 \times 1 \times 1$ domain resolved by a $64 \times 64 \times 64$ grid. The surface tensions are $\sigma_{lb} =0.15$, $\sigma_{ld} =0.4$ and $\sigma_{bd} =0.3$, giving a contact angle of $\theta_{l}=140.4^o$ and $Oh_d=0.058$.

\begin{figure}
\centering{
    \sidesubfloat[]{\label{drop2D3D:a}\includegraphics[scale=0.8]{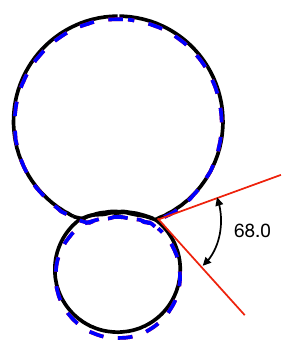}}
    \sidesubfloat[]{\label{drop2D3D:b}\includegraphics[scale=0.8]{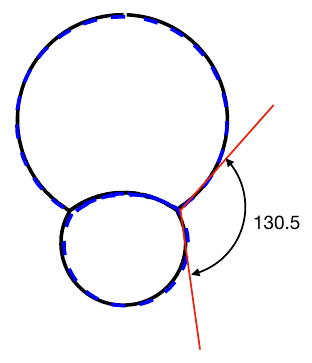}}}
\centering{
    \sidesubfloat[]{\label{drop2D3D:c}\includegraphics[scale=0.8]{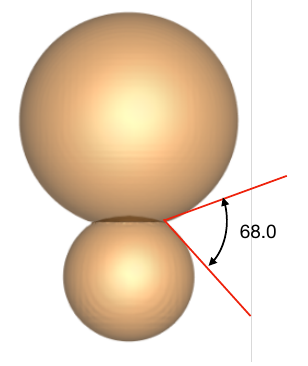}}
    \sidesubfloat[]{\label{drop2D3D:d}\includegraphics[scale=0.8]{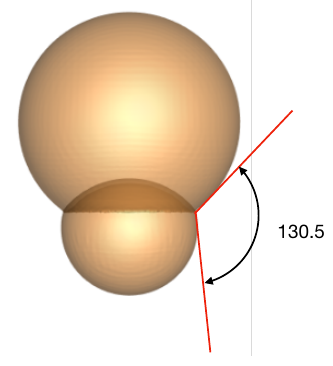}}}
\caption{The steady state for two-dimensional and three-dimensional flows using two different contact angles: (a, c) $\theta_l=68^o$ and (b, d) $\theta_l=130.5^o$. For the two-dimensional flows we show results for two grid resolutions.}
\label{drop2D3D}
\end{figure}

In Figure~\ref{drop2D3D} we examine the contact angles at equilibrium in more detail, for both two and three-dimensional flows.
The governing fluid parameters are $\rho_d=2$, $\mu_l=0.01$, $\mu_b=0.005$, and $\mu_d=0.02$. The bubble diameter is $d_b=0.5$ and the drop diameter is $d_d=0.3$. The surface tensions are $\sigma_{ld}=0.3$ and $\sigma_{bd}=0.35$ in Figure~\ref{drop2D3D}(\subref*{drop2D3D:a},\,\subref*{drop2D3D:c}), resulting in $\theta_l=68^o$, and $\sigma_{ld}=0.25$ and $\sigma_{bd}=0.2$ in Figure~\ref{drop2D3D}(\subref*{drop2D3D:b},\,\subref*{drop2D3D:d}), giving $\theta_l=130.5^o$. In Figure~\ref{drop2D3D}(\subref*{drop2D3D:a},\,\subref*{drop2D3D:b}), the solid lines are the results for a $64 \times 64$ uniform grid and the dashed lines are for a $32 \times 32$ uniform grid. 
We also plot two lines, forming a wedge with the right contact angle, showing that for both resolutions the computed angle agrees well with the theoretical prediction. 

\begin{figure}
\centering{
\sidesubfloat[]{\label{Engulfment1:a}
    \includegraphics[scale=0.325]{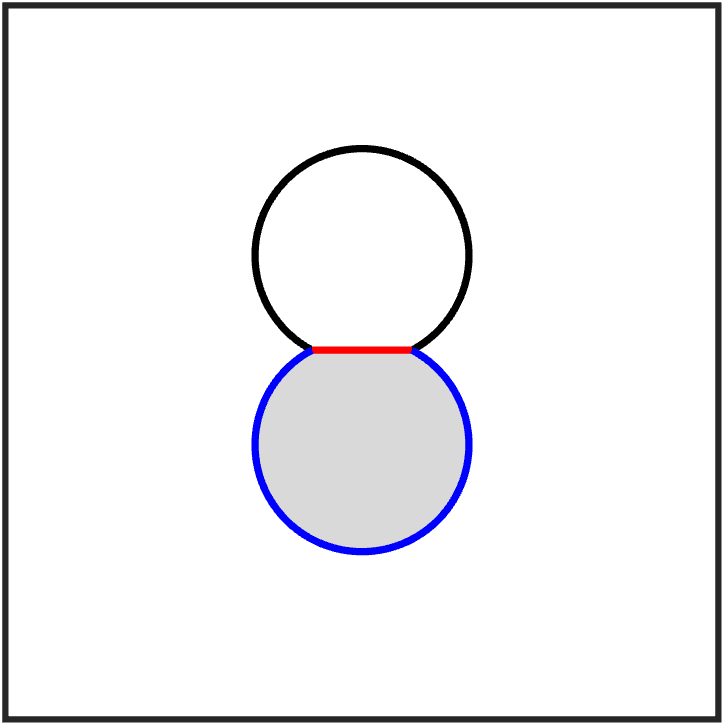}}
\ \ \
\sidesubfloat[]{\label{Engulfment1:b}
    \includegraphics[scale=0.325]{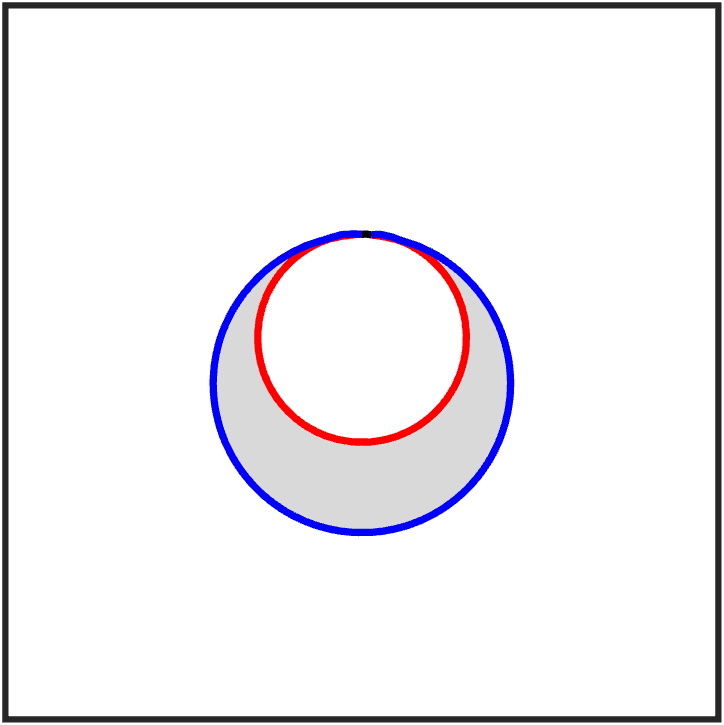}}
\ \ \
\sidesubfloat[]{\label{Engulfment1:c}
    \includegraphics[scale=0.325]{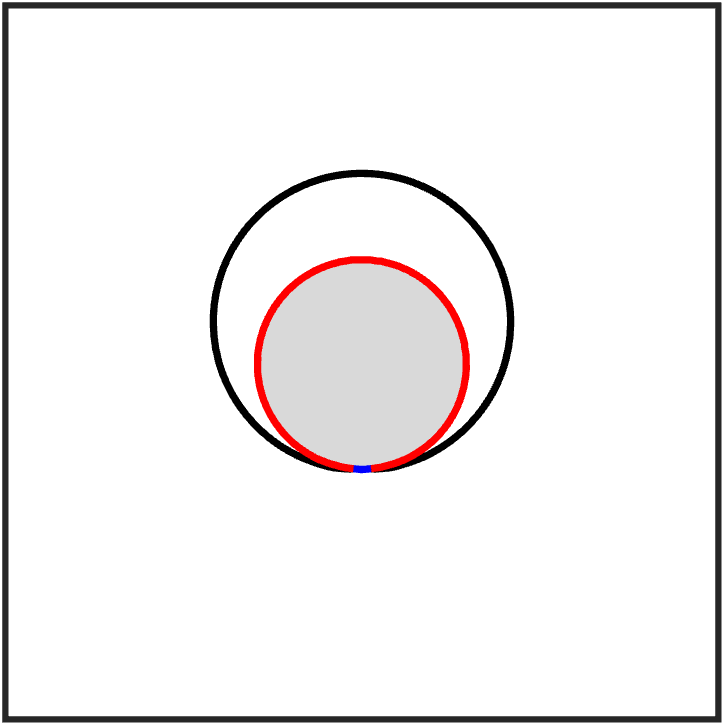}}}
\caption{Engulfment. (a) Initial conditions. (b) The steady state for a drop engulfing a bubble. (c)  The steady state for a bubble engulfing a drop.}
\label{Engulfment1}
\end{figure}

Figure~\ref{Engulfment1}\subref{Engulfment1:a} shows the initial state of two cases with an equal-sized bubble and drop where $d_b = d_d = 0.15$. Figure~\ref{Engulfment1}\subref{Engulfment1:b} shows a drop that nearly engulfs a bubble at the steady state and Figure~\ref{Engulfment1}\subref{Engulfment1:c} shows a bubble engulfing a drop. For drop engulfing the bubble, the surface tensions are $\sigma_{lb}=1.52$, $\sigma_{ld}=0.5$ and $\sigma_{bd}=1$, while for bubble engulfing the drop, the surface tensions are $\sigma_{lb}=0.5$, $\sigma_{ld}=1.52$ and $\sigma_{bd}=1$. All other parameters are the same as in Figure~\ref{dropresult}. Similar results have been presented by \cite{Zhaoetal2024}, where one phase fully contains the other phase. When we track the interface, we need to include topological changes, where we reconnect the fronts, to allow one phase to fully enclose the other one, and in Figure~\ref{Engulfment1} we have not done this, leaving a small segment of the interface between the bubble and the drop.

\begin{figure}
\centering{
\sidesubfloat[]{\label{ThreeDrops1:a}
    \includegraphics[scale=0.325]{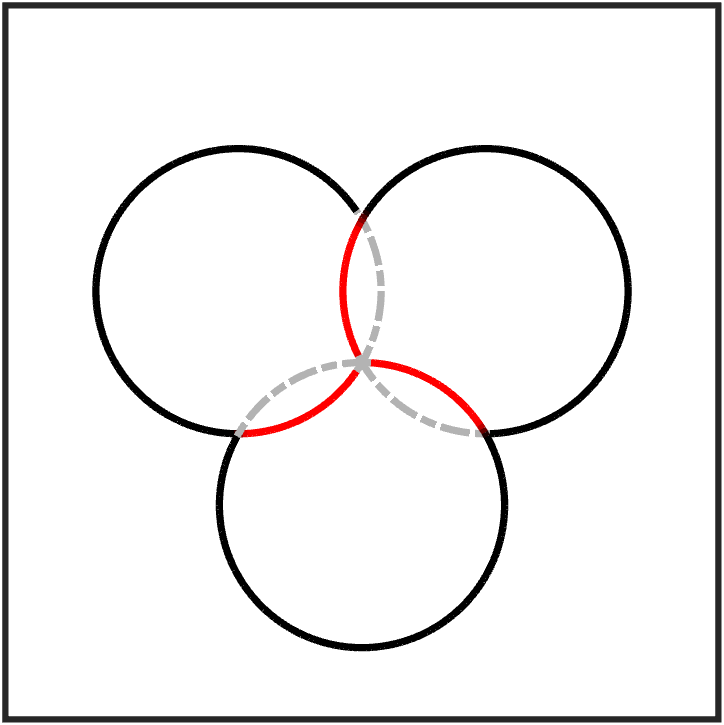}}
\ \ \
\sidesubfloat[]{\label{ThreeDrops1:b}
    \includegraphics[scale=0.325]{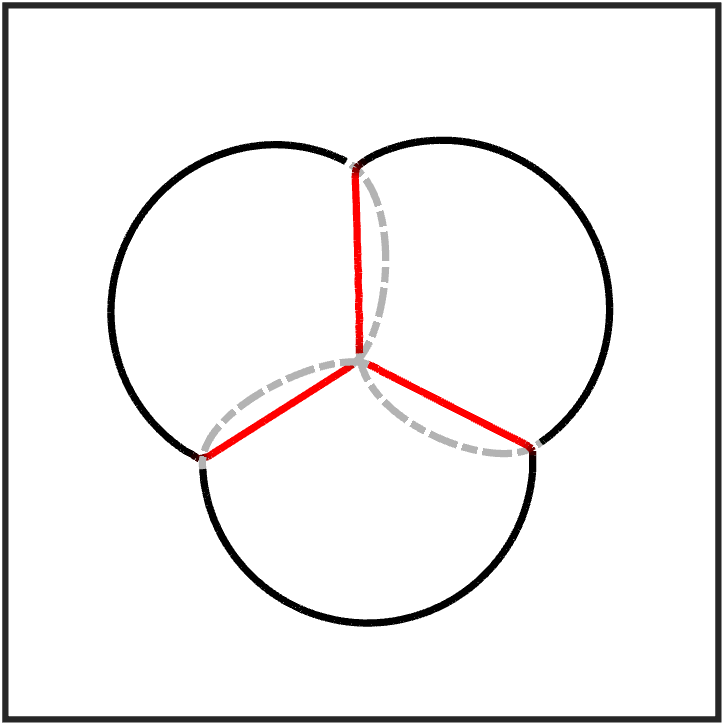}}
\ \ \
\sidesubfloat[]{\label{ThreeDrops1:c}
    \includegraphics[scale=0.325]{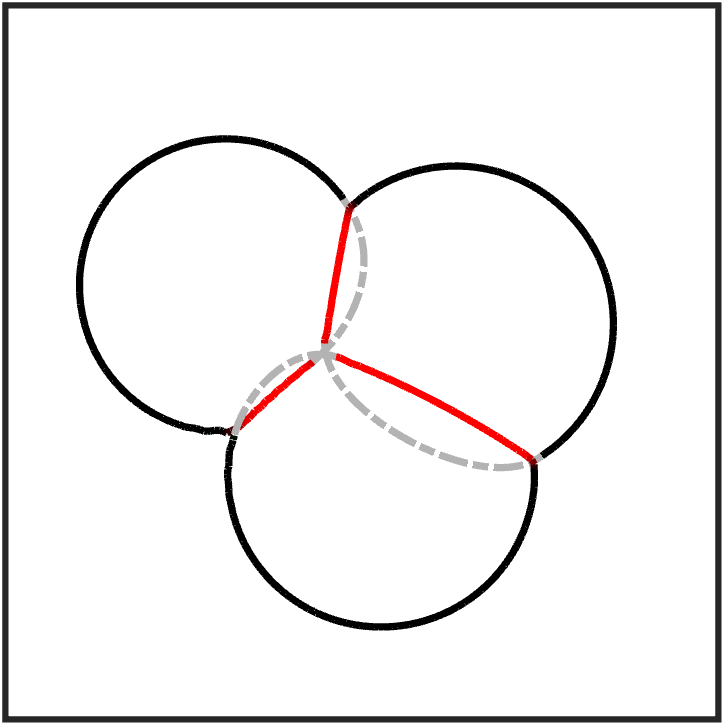}}}
\caption{The equilibrium configuration for three drops. (a) Initial conditions. (b) Drops with equal properties. (c) Drops with unequal properties.}
\label{ThreeDrops1}
\end{figure}

Using a virtual interface makes implementing relatively complex geometries involving triple points straightforward. \cite{Garckeetal2024} investigated a junction of three interfaces using a variational front tracking method, and here we used our untracked method for a similar setup, as shown in Figure~\ref{ThreeDrops1}. The initial configuration is shown in Figure~\ref{ThreeDrops1}\subref{ThreeDrops1:a} where the red lines show the physical interfaces between the drops and the gray dashed lines show the virtual interfaces. 
The flow is computed in a $1 \times 1$ domain using a $64 \times 64$ uniform grid.
In Figure~\ref{ThreeDrops1}\subref{ThreeDrops1:b}, we show the steady state results for drops with identical properties. The densities of the drops are $\rho_d=2$ and the density of the surrounding liquid is $\rho_l=1$. The viscosities are $\mu_d=0.04$ and $\mu_l=0.02$. The surface tension between the liquid and the drops is $\sigma_{ld}=0.1$, and the surface tensions between the drops are $\sigma_{12}=\sigma_{13}=\sigma_{23}=0.1$.
Figure~\ref{ThreeDrops1}\subref{ThreeDrops1:c} shows the steady state results with unequal properties. The densities for each drop are $\rho_{d,1}=0.5$ for the top-left drop, $\rho_{d,2}=1$ for the top-right drop, and $\rho_{d,3}=1.5$ for the bottom drop. The viscosities are $\mu_{d,1}=0.01$, $\mu_{d,2}=0.02$ and $\mu_{d,3}=0.03$, respectively. The surface tensions between the liquid and the drops vary, with $\sigma_{ld,1}=0.1$, $\sigma_{ld,2}=0.12$ and $\sigma_{ld,3}=0.14$. The drop-drop surface tensions are set as $\sigma_{12}=0.2$, $\sigma_{13}=0.17$ and $\sigma_{23}=0.14$. 

\subsection{Results for Two Fluids and a Solid}
\begin{figure}
\centering{
\sidesubfloat[]{\label{solidtransient1:a}
    \includegraphics[scale=0.275]{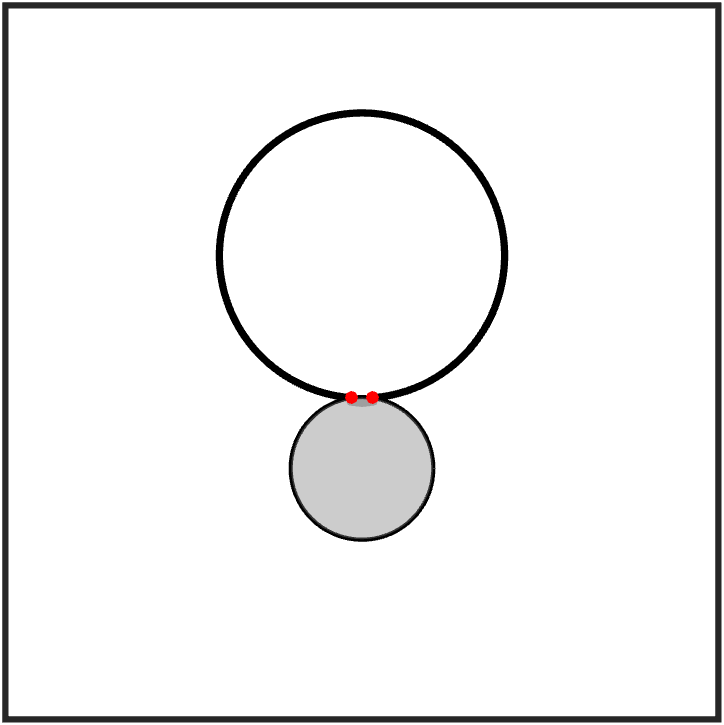}
    \includegraphics[scale=0.275]{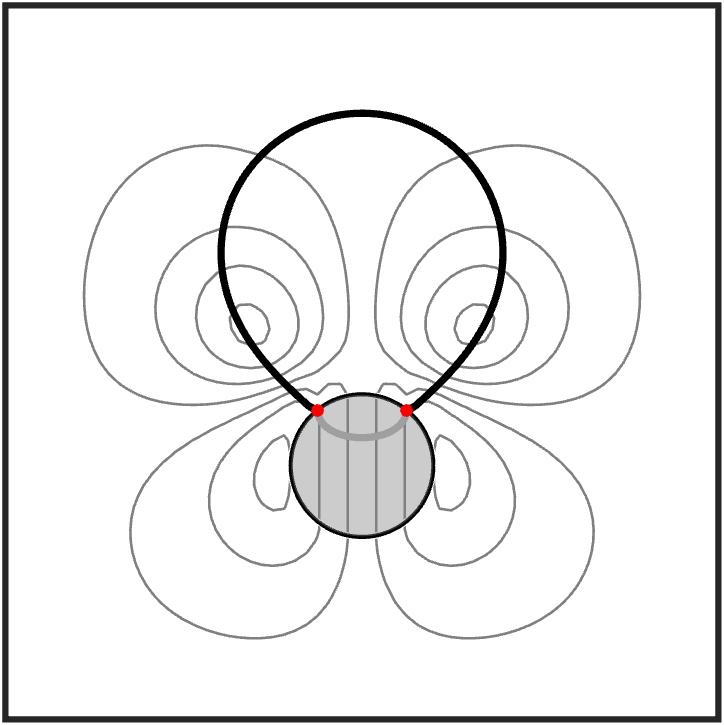}
    \includegraphics[scale=0.275]{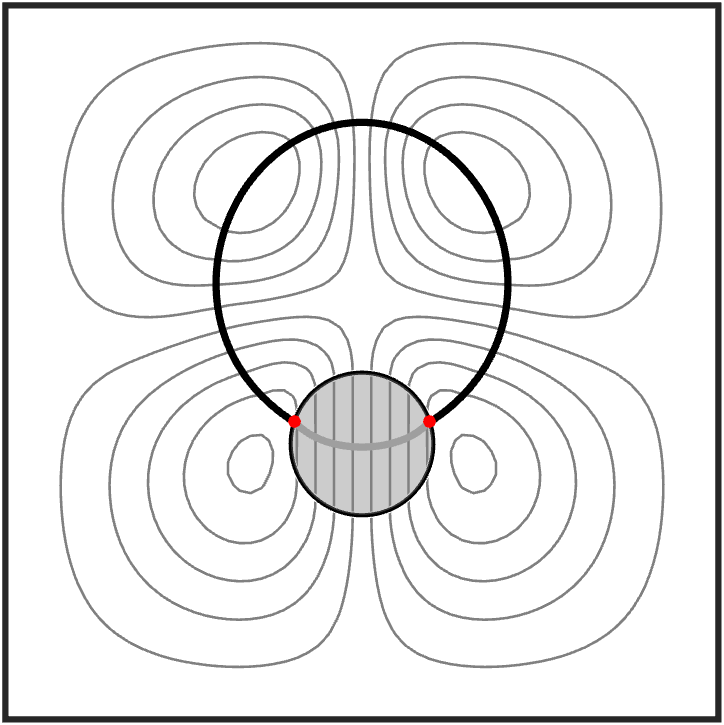}
    \includegraphics[scale=0.275]{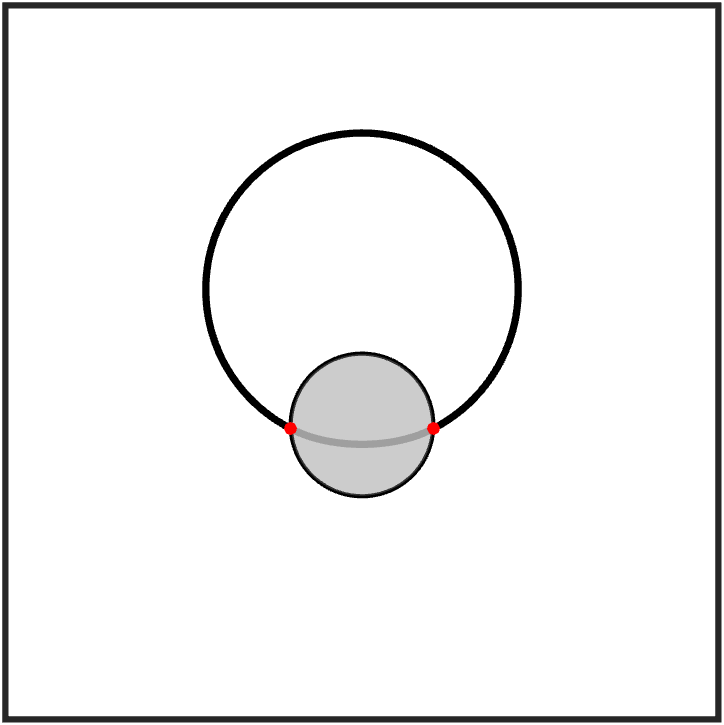}}
\vspace{.2cm}
\sidesubfloat[]{\label{solidtransient1:b}
    \includegraphics[scale=0.275]{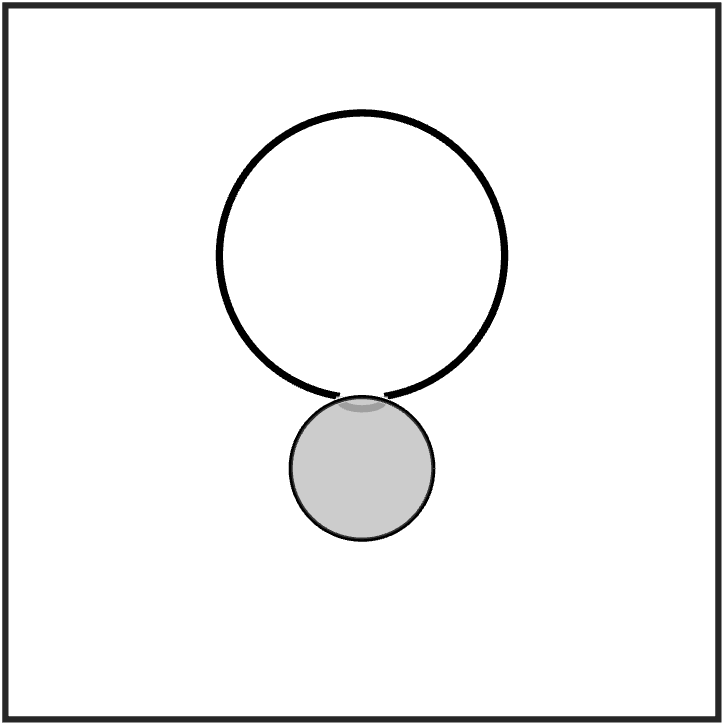}
    \includegraphics[scale=0.275]{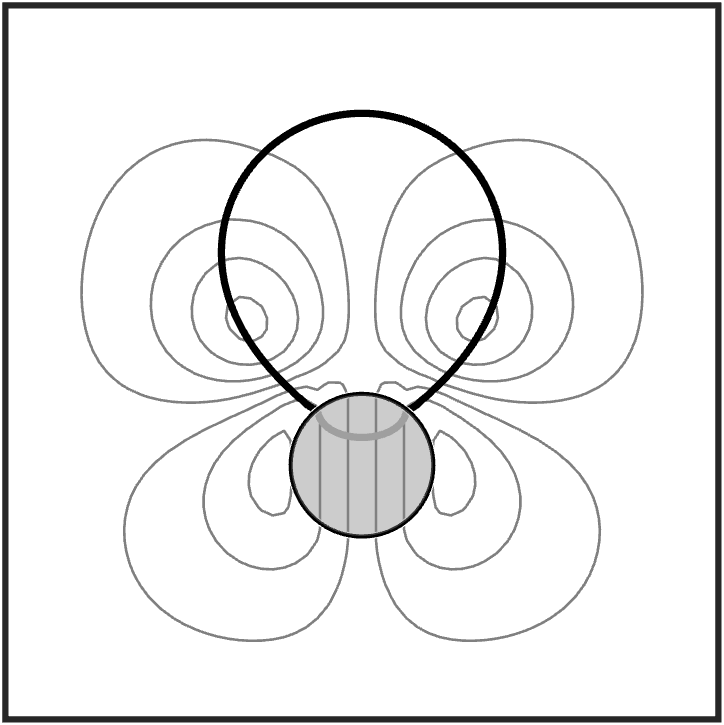}
    \includegraphics[scale=0.275]{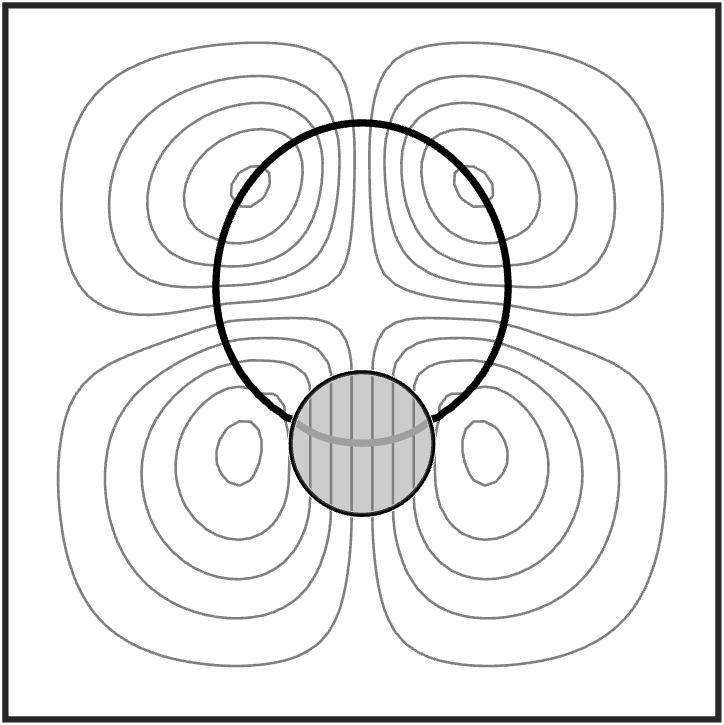}
    \includegraphics[scale=0.275]{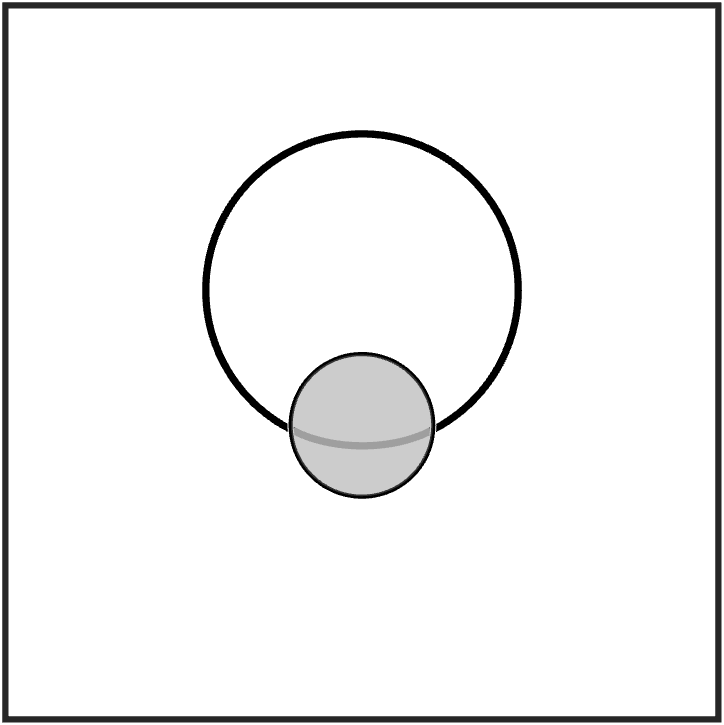}}}
\caption{Transient for a solid particle and a bubble at times 0, 0.2, 0.6 and 3.0. (a) Results for the tracked method. (b) Results for the untracked method.}
\label{solidtransient1}
\end{figure}

\begin{figure}
\centering{
\includegraphics[scale=0.45]{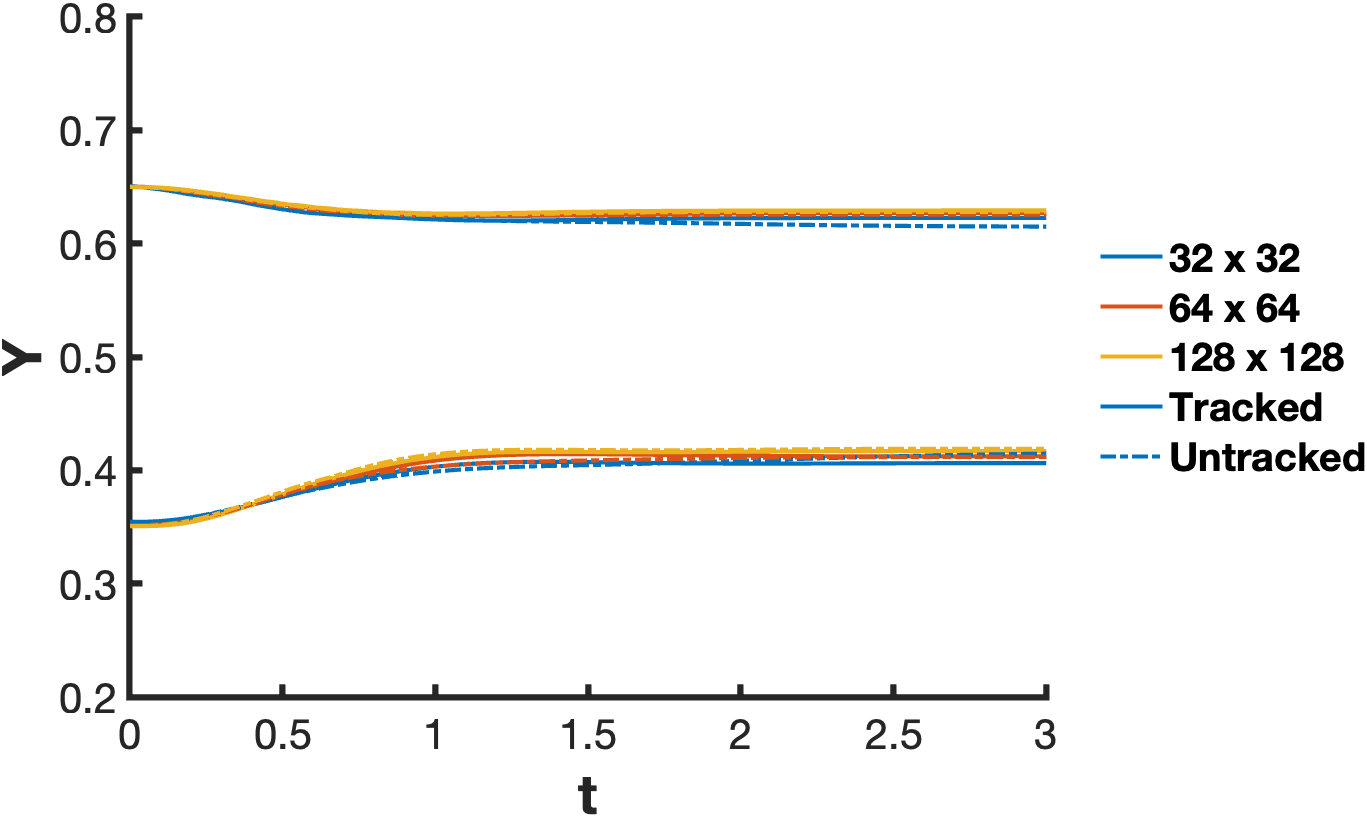}
}\caption{The motion of the centroids of the bubble and solid versus time for three resolutions and both approaches.}
\label{solidconvergence}
\end{figure}

 The engulfment of a solid particle (gray) into a bubble is shown in Figure~\ref{solidtransient1}. The physical and numerical parameters are the same as in Figure~\ref{dropresult}, but we have replaced the drop with the solid and use an equilibrium contact angle $\theta_o=120^o$ with a slip coefficient of $C_s=0.2$. The results computed by the tracked method are shown in Figure~\ref{solidtransient1}\subref{solidtransient1:a} and results computed using the untracked method in Figure~\ref{solidtransient1}\subref{solidtransient1:b}. We initially place the solid particle slightly inside the bubble and follow the motion until the contact angle has reached its equilibrium value. The initial conditions are shown in the first frame, and the subsequent frames show the interfaces and a few streamlines at times $t=0.2$ and $t=0.6$. The last frame shows the stationary equilibrium solution at time $t=3$. Figure~\ref{solidconvergence} shows the location of the centroids of the bubble and the solid versus time for three grid resolutions, using both the tracked and the untracked method. For the coarser grids we see a slight difference but as the grid becomes finer, the results all converge to the same answer.

\begin{figure}
\centering{
\includegraphics[scale=.45]{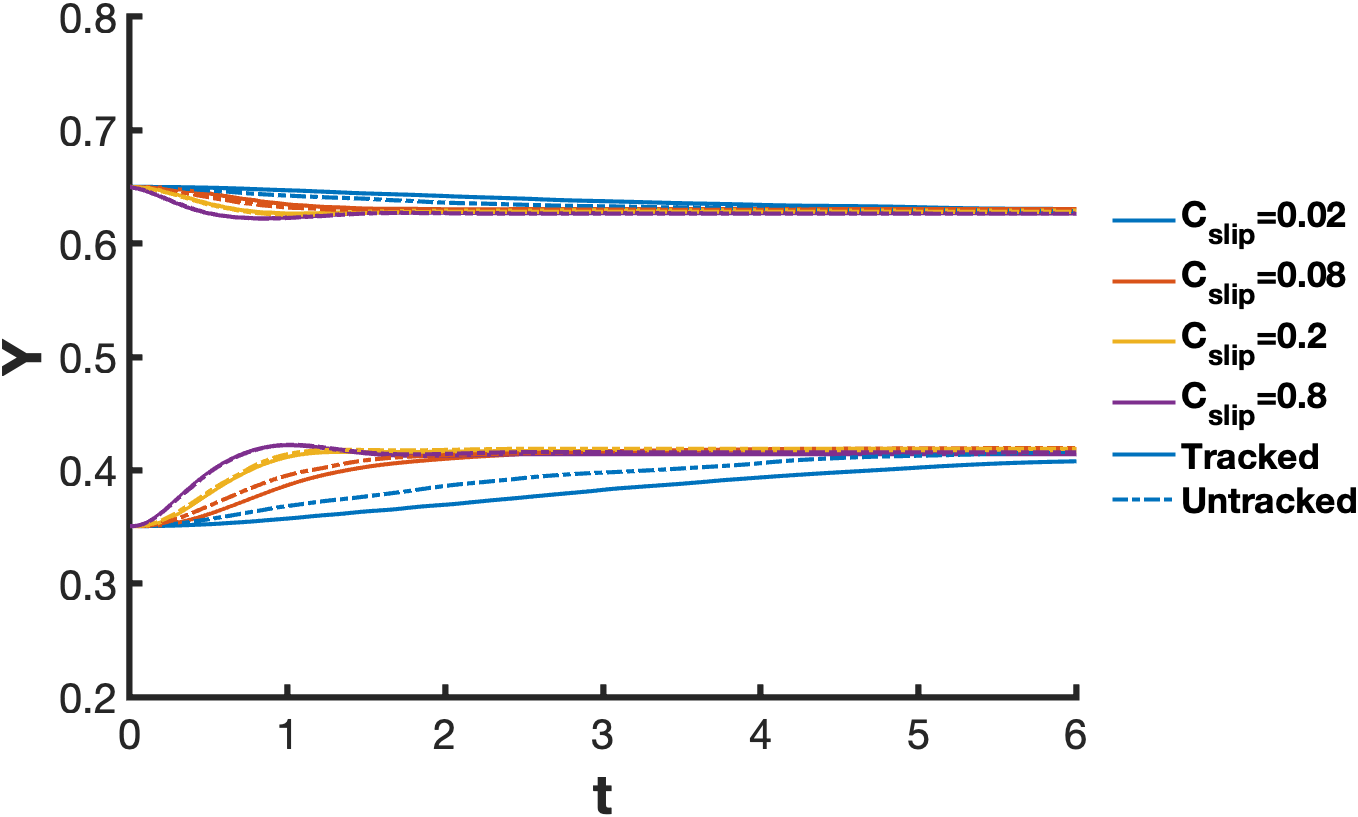}
}
\caption{The motion of the centroids of bubble and solid versus time for four different slip coefficients ($C_{slip}$)}
\label{Cslip}
\end{figure}

Although the slip velocity affects the speed of the contact line and thus the rate at which the solid moves into the bubble, it does not change the final equilibrium position. In Figure~\ref{Cslip}, we investigated four different slip coefficients using both the tracked and the untracked methods, maintaining the same properties as in Figure~\ref{solidtransient1}, using a grid resolution of $128 \times 128$. For the highest slip velocity the system reaches the equilibrium position quickly and both approaches give the same results, but for the lower slip coefficient where reaching a steady state takes longer time, the results for the tracked method lag the results for the untracked method slightly, for the resolution used here.

\begin{figure}
\centering{
\sidesubfloat[]{\label{solid3D:a}
    \includegraphics[scale=1.0]{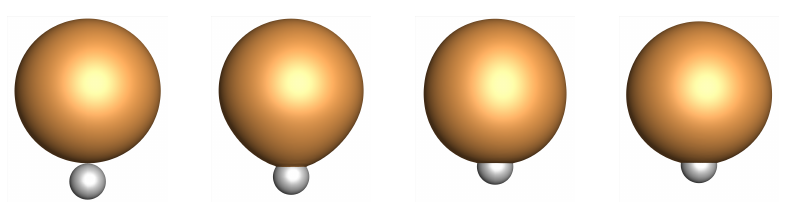}}}
\centering{
\sidesubfloat[]{\label{solid3D:b}
    \includegraphics[scale=1.0]{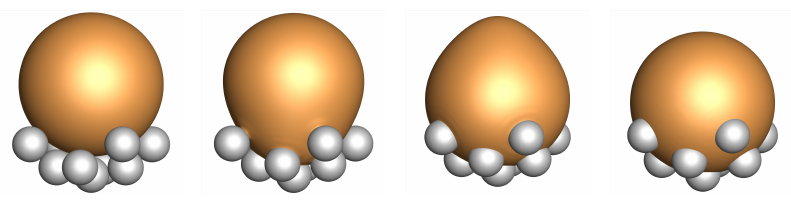}}}
\caption{Attachment of a bubble to (a) one solid particle and (b) nine particles. Time from left to right.}
\label{solid3D}
\end{figure}

The tracked method has been implemented in three dimensions. Figure~\ref{solid3D}\subref{solid3D:a} shows a single solid particle being captured, while Figure~\ref{solid3D}\subref{solid3D:b} shows the capture of nine particles randomly placed at the bottom of the bubble. The bubble has a diameter of $d_b=0.6$, and each solid particle has a diameter of $d_s=0.15$. The densities set for this simulation are $\rho_l=1.0$, $\rho_b=0.1$ and $\rho_s=2.0$, and the viscosities are $\mu_l=0.0025$, $\mu_b=0.00025$ and $\mu_s=0.0025$. The surface tension between the bubble and the liquid is $\sigma_{lb}=0.0324$ and the equilibrium contact angle is $\theta_o=100^o$. Both results are for a $1 \times 1 \times 1$ domain, resolved by a $64 \times 64 \times 64$ grid.

\begin{figure}
\centering{
\sidesubfloat[]{\label{dropsolid:a}
    \includegraphics[scale=0.275]{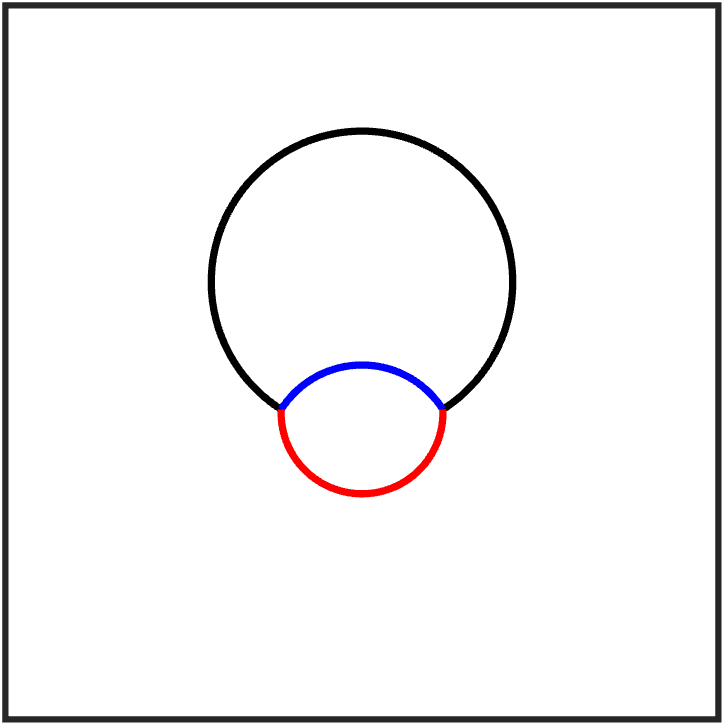}}
    \ \ \ \
\sidesubfloat[]{\label{dropsolid:b}
    \includegraphics[scale=0.275]{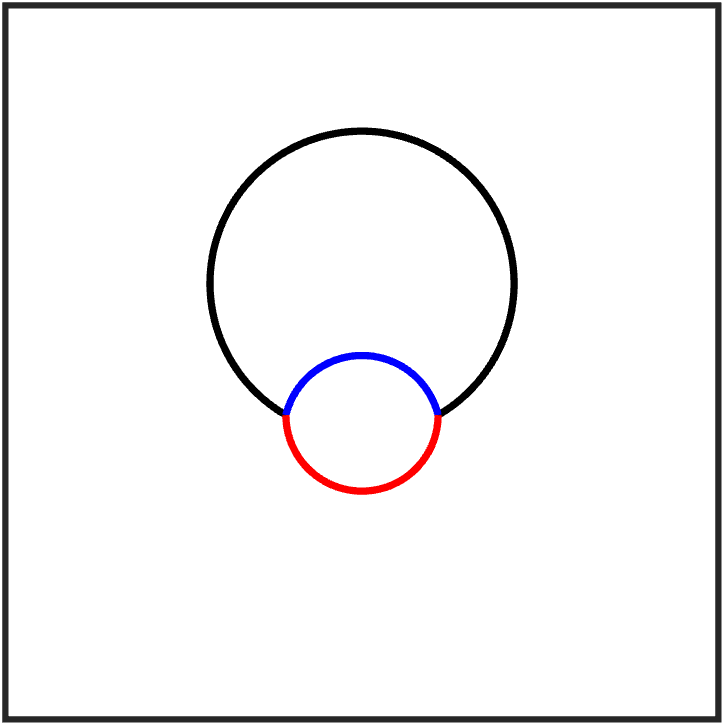}}}

\vspace{.2cm}
\centering{
\sidesubfloat[]{\label{dropsolid:c}
    \includegraphics[scale=0.275]{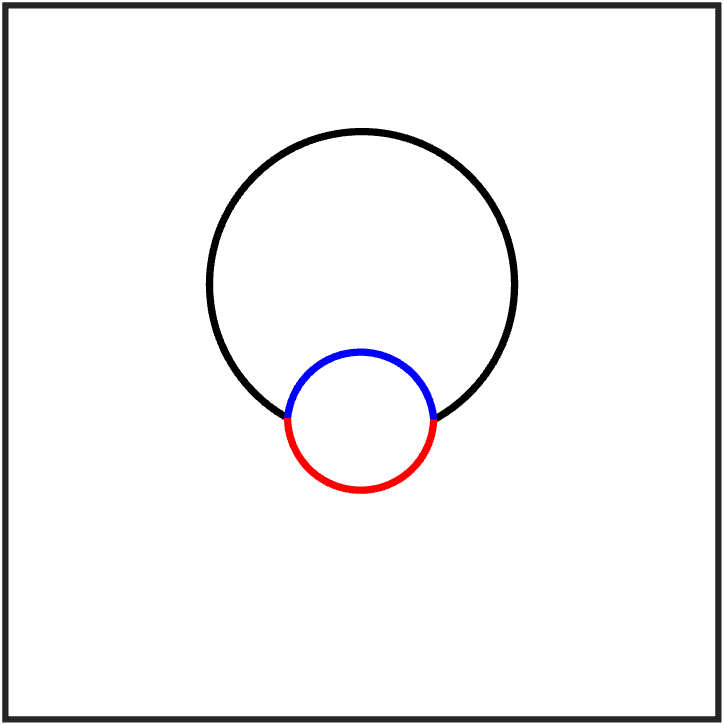}}
    \ \ \ \
\sidesubfloat[]{\label{dropsolid:d}
    \includegraphics[scale=0.275]{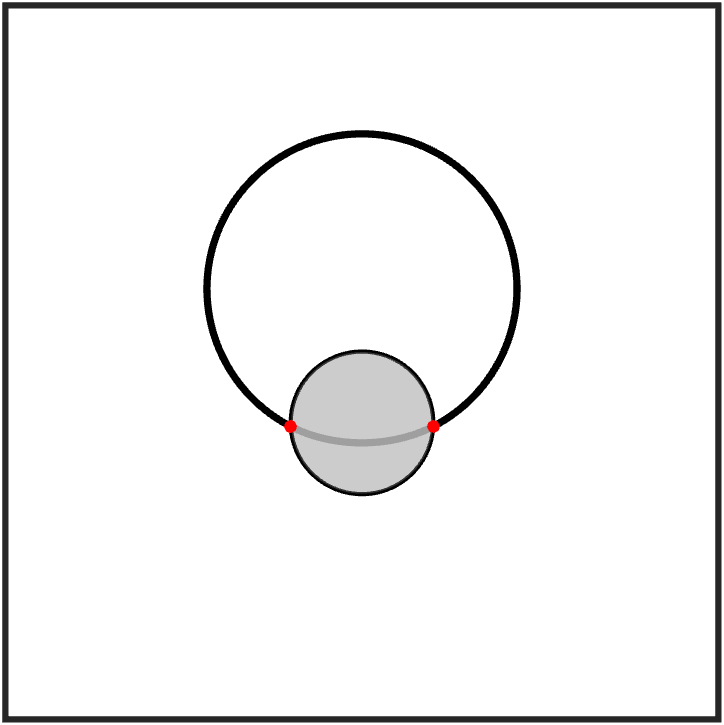}}}
\caption{Increasing the surface tension of a drop (a) $\sigma_{ld}=0.2$ ($\theta_d=150^o$), (b) $\sigma_{ld}=0.4$ ($\theta_d=166.1^o$), and (c) $\sigma_{ld}=0.8$ ($\theta_d=173.4^o$). (d) A solid particle (gray) attached to a bubble.}
\label{dropsolid}
\end{figure}

\begin{figure}
\centering{
\sidesubfloat[]{\label{dropsolid2:a}
    \includegraphics[scale=.3]{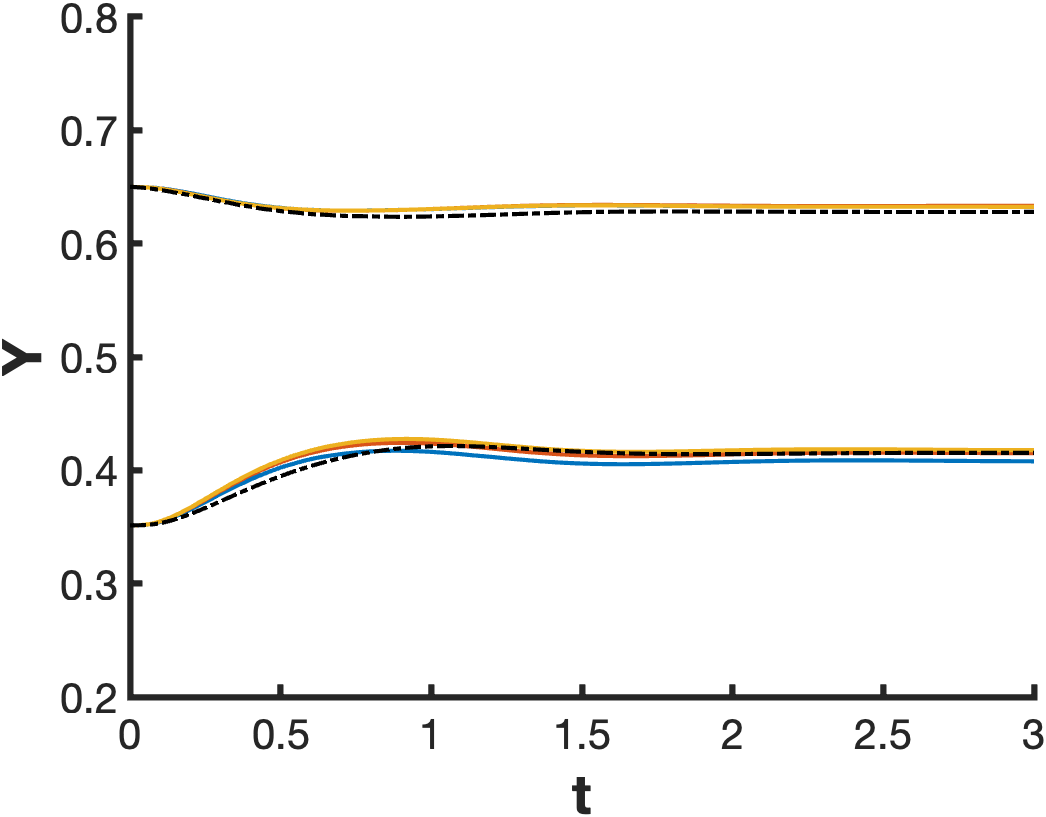}}
\ \ \
\sidesubfloat[]{\label{dropsolid2:b}
    \includegraphics[scale=.3]{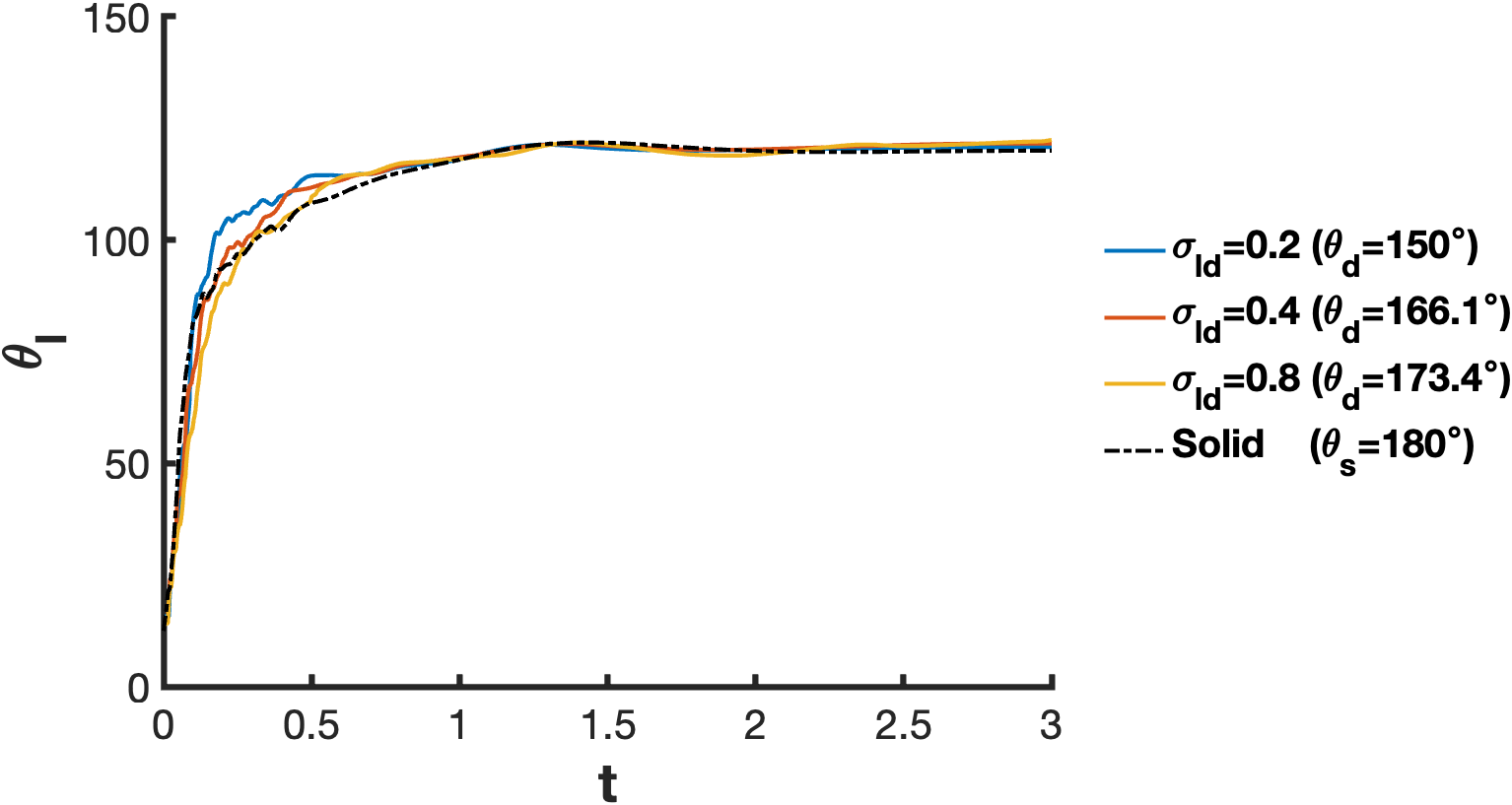}}}
\caption{(a) The movement of the centroids versus time for each phase. (b) The evolution of the liquid contact angle $\theta_l$ versus time for the drops and the solid.}
\label{dropsolid2}
\end{figure}

The major difference between a drop attached to a bubble and a solid sphere attached to a bubble is that the solid surface is locally flat at the contact line, whereas the drop surface is not. If the surface tension between the drop and both the liquid and the bubble is sufficiently large, we expect the equilibrium results for a drop to approach the solid results. In Figure~\ref{dropsolid}, this was tested by increasing the surface tension $\sigma_{ld}$ and $\sigma_{bd}$ while maintaining the equilibrium contact angle of $\theta_o=120^o$. The steady-state results for the drops with increasing surface tensions are: \subref{dropsolid:a} $\sigma_{ld}=0.2$ ($\theta_d=150^o$), \subref{dropsolid:b} $\sigma_{ld}=0.4$ ($\theta_d=166.1^o$), and \subref{dropsolid:c} $\sigma_{ld}=0.8$ ($\theta_d=173.4^o$). As a reference, Figure~\ref{dropsolid}\subref{drop2D3D:d} shows a solid attached to a bubble with slip coefficient $C_{slip}=0.5$. The simulations were conducted with the tracked method on a $128 \times 128$ grid, with other properties the same as in Figure~\ref{dropresult} for a drop and Figure~\ref{solidtransient1} for a solid. Figure~\ref{dropsolid2} compares \subref{dropsolid:a} the locations of the centroids of each phase and \subref{dropsolid:b} the evolution of the liquid angle $\theta_l$. The results indicate that with higher surface tension $\sigma_{ld}$,  drops increasingly mimic the behavior of a solid attached to a bubble.

\section{Conclusions}

We have described how to implement triple contact junctions, where three phases meet, into a front-tracking numerical method, for both three fluids systems and two fluids and freely moving solid particles. For solid particles, we use a strategy initially introduced by \cite{Fujitaetal2013, Fujitaetal2015} for level-set methods, where a fluid-fluid interface is extended into the solid as a virtual interface to help ensure that the solid experiences the correct force. The angle that the virtual interface makes with the fluid interface is equal to the equilibrium contact angle, and at equilibrium the interface extends smoothly through the solid surface. We explore two strategies for handling the contact point, one where we explicitly track it and another where we do not and treat it as a regular interface point. For the all-fluid problem, we also explore two strategies. In one we track the contact point separately but in the other we use a virtual interface with zero surface tension and treat the triple point as a regular interface point where the surface tension changes abruptly. Our results demonstrate that both the tracked and untracked approaches yield consistent and accurate predictions of the transient evolution and equilibrium configurations in various scenarios, including three fluids flows and the two fluids with a solid. The untracked method is slightly less accurate but simplifies the computational implementation significantly. We note that we have not addressed the time delay in the attachment, often seen when the liquid film between the two phases drains in a finite time. That is an important consideration since a drainage time that is longer than the time the particles stay in contact can prevent attachment. We hope to address that in a later publication.

\vskip 0.2in
\noindent
{\bf Acknowledgement}
\vskip 0.2in
This research was supported by NSF grant CBET-2035231. Computations were done at the Advanced Research Computing at Hopkins (ARCH) core facility  (rockfish.jhu.edu), which is supported by the National Science Foundation (NSF) grant number OAC1920103.

\section*{Appendix 1}
While solving equations (\ref{hbalance}) and (\ref{vbalance}) is straightforward, we include the steps here for completeness.
We start by defining
\begin{equation} 
 a=\frac{\sigma_{bd}}{\sigma_{lb}}; \quad  b=\frac{\sigma_{ld}}{\sigma_{lb}}; \quad c=\frac{\sigma_{ld}}{\sigma_{bd}}.
\label{angles1} 
\end{equation} 
The equations to be solved are therefore
\begin{equation} 
1+a \cos \theta_b +b \cos \theta_l =0, 
\label{equilibrium1}
\end{equation} 
\begin{equation}
a \sin \theta_b = b \sin \theta_l.
\label{equilibrium2}
\end{equation} 
The coefficients are not independent and, in particular, $b=ac$, or $c=b/a$.
We can, without a lack of generality, focus on finding $\theta_l$. 
Squaring equation (\ref{equilibrium2}), gives $a^2 \sin^2 \theta_b = b^2 \sin^2 \theta_l$, and using that $\sin^2 \theta=1-\cos^2 \theta$, yields
\begin{equation} 
a^2 (1-\cos^2  \theta_b) =b^2 (1-\cos^2 \theta_l),
\nonumber  \end{equation} 
or
\begin{equation} 
a^2 \cos^2  \theta_b=a^2-b^2 (1-\cos^2 \theta_l).
\nonumber  \end{equation} 
Equation (\ref{equilibrium1}) is then written
\begin{equation} 
a \cos \theta_b =-(1+b \cos \theta_l ) \quad \Rightarrow \quad 
a^2 \cos^2 \theta_b = (1+b \cos \theta_l )^2
\nonumber  \end{equation} 
Substituting for $\cos^2 \theta_b $ and expanding both sides gives
\begin{equation} 
a^2-b^2 +b^2 \cos^2 \theta_l = 1+2 b \cos \theta_l +b^2 \cos^2 \theta_l .
\nonumber  \end{equation}
Canceling the common terms yields 
\begin{equation} 
\cos \theta_l=(a^2 -b^2 -1)/ 2 b =\frac{1}{2} \Bigr( \frac{\sigma_{bd}}{\sigma_{ld}} \frac{\sigma_{bd}}{\sigma_{lb}}
-\frac{\sigma_{ld}}{\sigma_{lb}} -\frac{\sigma_{lb}}{\sigma_{ld}} \Bigr).
\label{generalsolution11}
\end{equation}
In the same way, we can solve for
\begin{equation} 
\cos \theta_b=(b^2 -a^2 -1)/ 2 a =\frac{1}{2} \Bigr( \frac{\sigma_{ld}}{\sigma_{bd}} \frac{\sigma_{ld}}{\sigma_{lb}}
-\frac{\sigma_{bd}}{\sigma_{lb}} -\frac{\sigma_{lb}}{\sigma_{bd}} \Bigr),
\label{generalsolution22}
\end{equation}
where we have substituted the definition of the coefficients $a$, $b$ and $c$. The third angle can be found by rotating the coordinate system or simply using the fact that the angles must add up to $2 \pi$.
\begin{equation} 
 \theta_b + \theta_d + \theta_l = 2 \pi
\label{sumofangles}
\end{equation} 

\section*{Appendix 2}
To generate a smooth virtual interface inside the solid, we minimize
\begin{equation}
E=\int \Big\vert \frac{\partial^2 {\bf x}}{\partial s^2} \Big\vert^2 ds,
\end{equation}
using an iteration.
We assume that $\Delta s$ is constant and can be left out of the minimization. The second derivative is approximated by
\begin{equation}
 \frac{\partial^2 {\bf x}}{\partial s^2} ={\bf x}_{l+1} -2 {\bf x}_{l} + {\bf x}_{l-1},
\end{equation}
and to minimize $E$ iteratively, we write the discrete terms that contain point $l$ 
\begin{equation}
E=\vert {\bf x}_{l} -2 {\bf x}_{l-1} + {\bf x}_{l-2} \vert^2 +
\vert {\bf x}_{l+1} -2 {\bf x}_{l}   + {\bf x}_{l-1} \vert^2 +
\vert {\bf x}_{l+2} -2 {\bf x}_{l+1} + {\bf x}_{l} \vert^2 
\end{equation}
and set the derivative of $E$ with respect to $x_l$ to zero:
\begin{equation}
\frac{\partial E}{\partial  x_l} =
  (x_l     -2 x_{l-1} + x_{l-2} ) 
-2(x_{l+1} -2 x_{l}   + x_{l-1} )+
  (x_{l+2}-2 x_{l+1}  + x_{l} ) = 0.
\end{equation}
Solving for $x_l$, we have:
\begin{equation}
x_l=\frac{1}{3} \Bigl[ 2 (x_{l+1}+x_{l-1} ) - \frac{1}{2} (x_{l+2}+x_{l-2} )\Bigr],
\label{App1-1}
\end{equation}
In practice, we iterate by a three-step process to avoid needing to involve neighbors more than one step away. 
First we rewrite equation (\ref{App1-1}) as
\begin{equation}
\begin{split}
x_l=\frac{1}{3} \Bigl[ 2 (x_{l+1}+x_{l}+x_{l-1} ) - \frac{1}{2} (x_{l+2}+x_{l+1}+x_{l}  ) \\
- \frac{1}{2} (x_{l}  +x_{l-1}+x_{l-2}) + \frac{1}{2} (x_{l+1}+x_{l}  +x_{l-1})
\Bigr] - \frac{1}{2}x_{l}.
\end{split}
\end{equation}
Then we define $\tilde x_l=(x_{l+1}+x_{l}+x_{l-1} )/3$ and write
\begin{equation}
x_l= 3 \tilde x_l-\frac{1}{2} ( \tilde x_{l+1}
+\tilde x_{l}+ \tilde x_{l-1} ) - \frac{1}{2}x_{l}.
\end{equation}
Moving the untilded term to the left hand side (thus under-relaxing) and defining ${\tilde {\tilde x}}_l=(\tilde x_{l+1}+\tilde x_{l}+\tilde x_{l-1} )/3$, we have $x_{l}= 2 \tilde x_l- {\tilde {\tilde x}}_l$. Similar equations are obtained for the other coordinates. 
One iteration step thus consists of:
\begin{equation}
[1]: \ \ \ \ \ \tilde {\bf x}_l= ({\bf x}_{l+1}+{\bf x}_{l}+{\bf x}_{l-1} )/3
\end{equation}
\begin{equation}
[2]: \ \ \ \ \ {\tilde {\tilde {\bf x}}}_l =(\tilde {\bf x}_{l+1}+\tilde {\bf x}_{l}+ \tilde {\bf x}_{l-1} )/ 3
\end{equation}
\begin{equation}
[3]: \ \ \ \ \ {\bf x}_l=2 \ \tilde {\bf x}_l -{\tilde {\tilde {\bf x}}}_l.  \ \ \ \ \ \ \ \ \ \ \ \ \ \ \ 
\end{equation}
We first find $\tilde {\bf x}$, then ${\tilde {\tilde {\bf x}}}$ and finally the new ${\bf x}_l$. Working only with the nearest points simplifies the programming, particularly for three-dimensional flows. In addition to making the iterations more compact, the rewrite results in under-relaxation that stabilizes the iteration.

\end{document}